\documentclass{aa}
\usepackage{graphicx}
\def\gtsima{$\; \buildrel > \over \sim \;$}
\def\ltsima{$\; \buildrel < \over \sim \;$}
\def\prosima{$\; \buildrel \propto \over \sim \;$}
\def\gsim{\lower.5ex\hbox{\gtsima}}
\def\lsim{\lower.5ex\hbox{\ltsima}}
\def\simgt{\lower.5ex\hbox{\gtsima}}
\def\simlt{\lower.5ex\hbox{\ltsima}}
\def\simpr{\lower.5ex\hbox{\prosima}}
\def\ml{$M_*/L$~}
\def\mlk{$M_*/L_K$~}
\def\mlr{$M_*/L_R$~}

\def\h1{$h^{-1}$}
\def\beq{\begin{equation}}
\def\eeq{\end{equation}}
\begin{document}
%\thesaurus{03(???)}
\title{
The K20 survey. VI. The Distribution of the  Stellar Masses in Galaxies up to $z\simeq 2$.
\thanks{Based on observations made at the European Southern Observatory,
Paranal, Chile (ESO LP 164.O-0560).}}
\author{
 	A. Fontana \inst{1}
	\and
	L. Pozzetti \inst{2}
	\and
	I. Donnarumma \inst {1}
	\and
	A. Renzini \inst{3}
	\and
	A. Cimatti \inst{4}
	\and
	G. Zamorani \inst{2}
	 \and
	N. Menci \inst{1}
	\and
	E. Daddi \inst{4}
	\and
	E. Giallongo \inst{1}
	\and
	M. Mignoli \inst{2}
	\and
        C. Perna \inst{1}
        \and
	S. Salimbeni \inst{1}
	\and
	P. Saracco \inst{5}
	\and
	T. Broadhurst \inst{6}
	\and
	S. Cristiani \inst{7}
	\and
	S. D'Odorico \inst{3}
	\and
	R. Gilmozzi \inst{3}
}

\institute{ 
INAF, Osservatorio Astronomico di Roma, via Frascati 33, Monteporzio, 00040, Italy
\and INAF, Osservatorio Astronomico di Bologna, via Ranzani 1, I-40127, Bologna, Italy
\and European Southern Observatory, Karl-Schwarzschild-Str. 2, D-85748, Garching, Germany
\and INAF, Osservatorio Astrofisico di Arcetri, Largo E. Fermi 5, I-50125 Firenze, Italy 
\and INAF, Osservatorio Astronomico di Brera, via E. Bianchi 46, Merate, Italy
\and Racah Institute for Physics, The Hebrew University, Jerusalem, 91904, Israel
\and INAF, Osservatorio Astronomico di Trieste, Via G.B. Tiepolo 11,
I-34131, Trieste, Italy
}
\offprints{Adriano Fontana, \email{fontana@mporzio.astro.it}}
\date{Received 3 November 2003; accepted 20 April 2004}

\abstract{We present a detailed analysis of the stellar mass
content of galaxies up to $z=2.5$ as obtained from the K20
spectrophotometric galaxy sample.  We have applied and compared two
different methods to estimate the stellar mass $M_*$ from broad--band
photometry: a Maximal Age approach, where we maximize the age of the
stellar population  to obtain the maximal mass
compatible with the observed $R-K$ color, and a Best Fit model, where
the best--fitting spectrum to the complete $UBVRIzJK_s$ multicolor
distribution is used. 
We find that the \ml~ ratio decreases with redshift: in particular,
the average \ml ratio of early type galaxies decreases with $z$, with
a scatter that is indicative of a range of star--formation time-scales
and redshift of formation.  More important, the typical \ml~ ratio of
massive early type galaxies is larger than that of less massive ones,
suggesting that their stellar population formed at higher $z$.
We show that the final K20 galaxy sample spans a range of stellar
masses from $M_*=10^9M_\odot$ to $M_*=10^{12}M_\odot$: massive
galaxies ($M_*\geq10^{11}M_\odot$) are common at $0.5<z<1$, and are
detected also up to $z\simeq 2$.  We compute the Galaxy Stellar Mass
Function at various z, of which we observe only a mild evolution
(i.e. by 20-30\%) up to $z\simeq 1$.  At $z>1$, the evolution in the
normalization of the GSMF appears to be much faster: at $z\simeq 2$,
about 35\% of the present day stellar mass in objects with $M_* \simeq
10^{11}M_\odot$ appear to have assembled.  We also detect a change in
the physical nature of the most massive galaxies: at $z \lsim 0.7$,
all galaxies with $M>10^{11}M_\odot$ are early type, while at higher
$z$ a population of massive star--forming galaxies progressively
appears.
We finally analyze our results in the framework of $\Lambda$--CDM
hierarchical models.  First, we show that the large number of massive
galaxies detected at high $z$ does not violate any fundamental
$\Lambda$--CDM constraint based on the number of massive DM
halos. Then, we compare our results with the predictions of several
renditions of both semianalytic as well as hydro-dynamical models.
The predictions from these models range from severe underestimates to
slight overestimates of the observed mass density at $\leq 2$.  We
discuss how the differences among these models are due to the
different implementation of the main physical processes.
 }
\titlerunning{The K20 Galaxy Stellar Mass Function}
\authorrunning{A. Fontana et al.}  \maketitle

\section{Introduction}

The recent consolidation of the ``concordance'' cosmological scenario
(Bennett et al. 2003), where several independent observational
evidences have provided precise measures for the basic cosmological
parameters, is opening a unique opportunity to understand the
processes that led to galaxy formation and evolution. Without much
residual ambiguity about the redshift-cosmic time relation and the
dark energy/dark matter content of the universe, observations of
galaxies at low and high redshift can better shed light on such
processes as a function of both cosmological time and local
over-density.

In the ``concordance'' cosmological scenario, the history of galaxies
is driven by the build-up of the stellar population contained in their
dark matter halos.  Hierarchical theories of galaxy formation are
characterized by a gradual enrichment of the star content of galaxies
as a result of gas cooling within dark matter halos, and of
progressive growth of the galaxy mass through merging events which may
also promote massive star-bursts. However, different renditions of the
hierarchical paradigm can differ dramatically in their predictions. In
some cases an extremely rapid decrease of the number density of massive
galaxies with increasing redshift is predicted (e.g., Baugh et
al. 2003), while in other cases such decrease does not start until
beyond $z\sim 1$ (e.g., Nagamine et al. 2001a,b; Hernquist \& Springel
2003; Somerville et al.  2004a, Nagamine et al. 2004). Clearly, the
direct mapping of galaxy evolution through cosmic time can effectively
restrict the choice among such models.

Within this framework, $K$--band surveys have long been recognized as
ideal tools to study the process of mass assembly at high redshift
(Broadhurst et al. 1992, Gavazzi et al. 1996; Madau, Pozzetti \&
Dickinson 1998).  With respect to optical bands, indeed, the $K$ band
samples up to high $z$ the rest frame optical and near--IR spectral
range, and therefore it is less sensitive to the instantaneous star
formation activity and to dust extinction. Albeit the relation between
near--IR luminosity and stellar mass is not univocal, deep imaging and
spectroscopic surveys have been  carried on to
test the cosmological scenarios on mass--selected galaxy samples
(Songalia et al. 1994; Kauffmann \& Charlot 1998;
Fontana et al. 1999; Cohen et al. 1999; \cite{drory}; Firth et al. 2002).

The K20 survey (Cimatti et al. 2002a) has been designed to extend and
complement these studies, with the explicit aim of investigating the
high redshift evolution of massive galaxies. It is based on a sample
of about 500 galaxies to $K_s<20$, for which a nearly complete
spectroscopic identification and a deep $UBVRIzJK_s$ multicolor coverage
is available, which together make it an ideal dataset to study the
evolution of a mass-selected sample of galaxies up to $z\simeq 2$.  In
the K20 dataset, the evolution of bright, massive galaxies has been
investigated up to $z\simeq 2$ through the study of the $K$--limited
redshift distribution (Cimatti et al. 2002b) and the near--IR luminosity
functions (\cite{pozzetti2003}). The results of the K20 survey show that
galaxies selected in the $K$ band are characterized by a modest luminosity
evolution up to $z\simeq 1$, that seems well  described by simple pure
luminosity evolution (PLE) models.

In this paper, we will use the K20 dataset to directly study the
evolution of the stellar mass content in galaxies up to $z\simeq 2$.
Recently, various techniques have been developed to directly estimate
the stellar mass content of galaxies up to $z\sim 3$. Some rely on
detailed spectral analysis (Kauffmann et al. 2003, for low $z$
galaxies), others on multi-wavelength imaging observations to remove
or reduce the uncertainties involved in the conversion between
near--IR luminosity and stellar mass (Giallongo et al. 1998;
Brinchmann \& Ellis 2000; Cole et al. 2001; Papovich et al. 2001;
Shapley et al. 2001; Drory et al. 2001; Dickinson et al. 2003 - D03
hereafter-; Fontana et al. 2003 - F03 hereafter-; Rudnick et al. 2003;
Saracco et al. 2004).

The applications of such techniques have been primarily driven by the
available datasets.  On a relatively shallow dataset, Giallongo et al.
(1998) emphasized that at $z\simeq 0.7$ blue ``faint'' galaxies are an
order of magnitude less massive than redder galaxies of comparable $B$
magnitude, and attempted a first estimate of the evolution of the
cosmological mass density up to high $z$.  Brinchmann \& Ellis (2000)
used a multi-wavelength coverage of the CFRS galaxy sample (Lilly et
al. 1995) to trace the stellar mass density in morphologically
selected galaxy samples up to $z=1$, while \cite{drory} used a larger
set of galaxies with photometric redshifts to set an upper limit on
their mass density at $z\simeq 1$. Papovich et al. (2001) and Shapley
et al. (2001) pushed this technique to its limit to constrain the mass
of Lyman--break galaxies, showing that objects with stellar mass in
excess of $10^{10} M_\odot$ are commonly detected in the $z\simeq 3$
universe. Recently, D03, F03 and
Rudnick et al. (2003) used extremely deep optical and near--IR data in
the HDF--N and HDF-S to derive the evolution of the
stellar mass density from $z=0$ to $z=3$.  These studies have
derived a fast evolution of the stellar mass density in the redshift
range $z=1-3$, but have also shown that large ambiguities still
persist due to cosmic variance and uncertainties due to incomplete
spectroscopic redshift coverage.

The plan of the paper is as follows: in Section 2 we describe the data
sample used in this paper. In Section 3 we discuss and compare the two
methods we have applied to estimate the stellar masses of the K20
galaxies.  Further details and validation tests of these procedures
are deferred to Appendix A for the interested reader.  In Section 4,
we present the derived evolution of stellar masses and rest-frame \ml
ratios. In Section 5, we present the observed Galaxy Stellar Mass
Functions and global Mass Density, for the total sample and for
different spectral types, while a more technical discussion of the
corrections required to take into account the effects of
incompleteness is presented in Appendix B.  In Section 6 we compare
our results at different redshifts with the prediction of different
renditions of the CDM models for galaxy formation.  Finally, in
Section 7 we summarize the results and discuss their validity and
implications in the general scenario of galaxy formation.

A Salpeter IMF and the ``concordance'' cosmology ($H_0=70$ km s$^{-1}$
Mpc$^{-1}$, $\Omega_m=0.3$ and $\Omega_{\Lambda}=0.7$) are adopted
throughout the paper. In the following, we shall often refer 
to the \ml~ ratio, that is always computed in solar units adopting
$K_\odot=3.36$ and $R_\odot = 4.48$

\section{The Data}

The K20 sample (Cimatti et al. 2002a) has been selected at $K_s<20$
(Vega system) over two independent fields, namely a sub--area of the
Chandra Deep Field South and the field around the $z=3.7$ QSO
Q0055-269, for a total of 546 objects over an area of 52 arcmin$^{2}$.
Spectra have been obtained with the ESO-VLT, mostly using the optical
spectro--imagers FORS1 and FORS2, with the addition of a few redshifts
obtained in the IR with ISAAC. Besides the spectra already presented
in Cimatti et al. (2002a), we use here additional spectra recently
obtained with the same instruments, partly described in \cite{daddi04}
and Cimatti et al. (2003), and partly obtained within the ESO public
follow-up of the GOODS survey (Vanzella et al. 2004, in preparation).

In addition to the spectroscopic observations, we have used deep
multicolor coverage ($UBVRIzJK_s$) of the whole galaxy sample obtained
from targeted and public observations with VLT and NTT made available
through the ESO Archive. From such detailed color information, we have
derived well calibrated photometric redshifts for the whole galaxy
sample, including those objects for which no spectroscopic redshift
could be obtained.  The resulting photometric redshift dispersion is
$\sigma(z_{\rm spec}-z_{\rm phot})/(1+z_{\rm spec})= 0.05$ (Cimatti et
al. 2002a).

The K20 redshift survey is at present the most complete $K$--selected
deep spectroscopic survey so far obtained. The spectroscopic coverage
on the total sample is 92\%, and excluding spectroscopically confirmed
stars and AGNs, the total galaxy sample with either spectroscopic or
accurate photometric redshift is made of 487 objects, 446 of which
(i.e. 92\%) have a spectroscopic redshift.   Multicolor imaging in the
$UBVRIzJK$ bands is also available for all galaxies, with the
exception of three galaxies in the CDFS field where only $R-K$ is
available.

In the following we shall also make use of a simple spectroscopic
classification: at $z\leq 1.6$, we have named ``early type'' galaxies
as objects identified by absorption lines and with no detected
emission lines; ``early+emission type'' galaxies, similar to early
type but with a weak [OII]$\lambda$3727 emission line; ``late type'' galaxies,
i.e. star--forming objects with a strong [OII]$\lambda$3727  emission line.
At $z>1.6$, where the [OII]$\lambda$3727  doublet is not observable, we
have classified three objects in the CDFS as early type, since
their spectra are dominated by features of evolved stellar
populations, and the others as late type since they are UV-bright
galaxies with far--UV absorption lines.  Overall, the K20 sample
includes 107 early type, 44 early+emission type and 297 late type
galaxies.

In this study we have considered the whole K20 galaxy sample,
i.e. including also the galaxies located in the large structures at
$z=0.65-0.73$ in both the Q0055 and CDFS fields.  Based on the number
and spatial distributions of members, on the X-ray luminosity and B
luminosity of the bright central elliptical, we indeed estimate that
at most one of these structures can be classified as a richness 0
cluster in the Abell classification scheme, while the others are
likely to be associated to either poorer clusters (e.g. groups) or
loose, extended, sheet-like structures (see Gilli et al. 2003 for a
discussion of the K20 structures in the CDFS field).  Based on recent
cluster catalogs we estimate that the number of clusters with Abell
richness $\ge 0$ over the K20 area is of the order of 0.7 (see, for
example, Table 6 in Postman et al. 2002), similar to what predicted by
numerical simulations (Evrard et al. 2002).  On the basis of this
comparison we conclude that the number of redshift peaks in our data
is in fair agreement with both existing data from large areas and
theoretical simulations, and we shall therefore use the whole K20
sample when computing mass functions and other integrated quantities.

We finally note that all the magnitudes of the K20 sample are
estimated in ``optimal'' Kron apertures, as obtained from the
SExtractor package (Bertin \& Arnouts, 1996). ``Kron'' magnitudes are
prone to systematic underestimates of the total flux, by an amount
that depends on the morphology, sampling, redshift and S/N of the
objects. With detailed simulations we have shown that at $z\sim 1$ the
underestimate is negligible for compact objects, and is about 10\% for
spirals and 20\% or even more for $L>L_*$ ellipticals
(\cite{cimatti02k20}).
We have decided {\it not} to apply any correction in the final 
stellar mass distributions,
since the morphological mix is expected to change significantly across
the mass and redshift range that we sample, but we shall  discuss
the effects of an average correction of $\simeq 20\%$ when comparing
our data with the theoretical predictions, as it has already been done
in our previous papers (e.g., Cimatti et al. 2002b, Pozzetti et al. 2003).

\section{The estimate of the galaxy stellar masses}

It is widely acknowledged that a close relationship exists between the
near--IR light and the stellar mass in local galaxies (e.g. Gavazzi et
al. 1996; Madau, Pozzetti \& Dickinson 1998). However, an accurate
estimate of the galaxy stellar mass at high $z$, when galaxies are
observed at various evolutionary stages, is more uncertain because of
the variation of the $M_*/L_K$ ratio as a function of the age
and other parameters of the stellar population.
The typical $M_*/L_K$ ratios for  exponentially declining star
formation histories range from small values ($M_*/L_K\leq 0.1$) at
young stellar ages $\leq 0.1$~Gyr to values about unity after 10
Gyrs. This variation in the \mlk \ ratio is larger than the scatter
induced (at a given age) by different metallicities or star formation
time-scales.
In addition, in the case of real galaxies the possibly complex star--formation
histories and in particular the presence of minor bursts of star
formation can affect the derived $M_*/L_K$ and therefore their mass estimate.

Additional information on the spectral energy distribution of
individual galaxies, as the multiband imaging available in our sample,
can be used to overcome or at least minimize this problem.
The use of the rest--frame optical and UV bands is a way to correct
(at least conceptually) for the contribution of high-luminosity and
low-mass young stellar population to the observed IR light.

Even with these additional data, however, the actual star--formation
history of individual galaxies cannot be unambiguously recovered, and
we are forced to rely on simplifying assumptions on the plausible
star--formation histories.  We will apply and compare here two
different methods to estimate the stellar masses from the observed
magnitudes, that are based on different assumptions on the previous
star--formation history.
For both we will adopt the Bruzual and Charlot (2003) code for
spectral synthesis models, in its more recent rendition, using its low
resolution version with the ``Padova 1994'' tracks.  We have
investigated about possible systematic differences between this recent
version and the previous one (GISSEL 2000) used in previous similar
works we will compare with (D03, F03).
We have found that the spectra obtained with the new version
are remarkably similar to the previous ones, such that the integrated
magnitudes are similar to a few hundreds of magnitude. As a result,
the estimated stellar masses result comfortably similar to the BC00
ones, with an average offset of only $-0.04dex$ and a scatter of
$0.12dex$.  In the following, we shall therefore compare our results
with those of previous surveys without any further re-normalization.

In our analysis we will adopt only the classical Salpeter IMF
(Salpeter 1955). The same IMF has been used in several previous works
that we shall compare with (e.g. Brinchmann \& Ellis 2000; Cole et
al. 2001; D03; F03) as well as in
some of the theoretical predictions that will be tested with our
data. Unfortunately, scaling to different IMFs may not be simple, since
these corrections depend on the age of the stellar population. For
instance, for the simple case of simple stellar populations, the \ml \
ratio of a Salpeter IMF at ages $<1$~Gyr is larger by a factor 1.3-1.5
with respect to the case of the frequently used Kennicutt IMF
(Kennicutt 1983), and by a factor up to $\sim 2.2$ at larger ages
(this factors are largely independent of the wavelength).
With respect to the Kroupa IMF (Kroupa 2001), the \ml \ ratio from a
Salpeter IMF is systematically larger by a factor $\sim 1.6$, roughly
independent of the age of the population.

\subsection{The ``Maximal Age'' method}

A first approach that we follow is to {\it assume a simple scenario}
for the star--formation history of the different types of
galaxies. This can be done adopting a limited set of evolutionary
models, chosen in order to reproduce the colors of local galaxies,
following the spirit of PLE models (e.g. \cite{pozzetti2003}) with a
fixed redshift of formation.

In our case, we have adopted the parameterization used by Cole et
al. (2001), a choice which is particularly useful for comparison with
their local GSMF. It consists of a set of models with exponentially
declining star formation rate, with time-scale $\tau \geq 1$ Gyr,
metallicities ranging from 0.2$Z_\odot$ to 2.5$Z_\odot$, a constant
dust absorption taken from the Ferrara et al. (1999) model, all
computed assuming that the star-formation history of each galaxy
started at $z_{\rm form}=20$. The time-scale of star formation is then
determined for each galaxy by demanding to the model to reproduce the
observed $R-K_s$ color, which is more sensitive to the spectral type
than the $J-K_s$ color used by Cole et al. (2001).The mass is then
derived by normalizing the model to the observed $K$-band luminosity.

We note that this approach maximizes the {\it age} of the stellar
population (hence its \ml \ ratio) as much as possible within the
current CMB constraints.  To emphasize this aspect, in the following
we refer to these models as ``Maximal Age'' (MA) models.  We note that
this choice is less extreme than the ``Maximal Mass'' model of
\cite{drory}, that fits the observed $K$ band only assuming
redshift--dependent maximal $M_*/L_K$, in practice ignoring the
contribution of star--forming populations to the observed $M_*/L_K$
ratio.

We have verified that the resulting values of the stellar mass are not
very sensitive to the particular set of parameters adopted in these
models. In particular, we have also explored the dust--free
`PLE-like'' star--formation histories of Pozzetti et al. (2003) with
$z_{\rm form}\sim 6$ and solar metallicity, finding that the resulting
stellar masses in our sample are quite similar to those of the Cole et
al. (2001) parameterization that we have adopted.

Adopting other colors from any of all the available bands %
we have found that the resulting masses do not vary by more than 5\%
(r.m.s.) on average.

\subsection{The ``Best Fit'' method}

\subsubsection{The method}

In an effort to release as much as possible the underlying assumptions
on star formation histories and other galaxy properties, we have also
used a larger set of galaxy templates, spanning a much wider parameter
space, and applied it to the full multicolor spectral energy
distribution (SED) to constrain their allowed range for each object. This
technique has already been widely applied in previous studies (e.g.,
Giallongo et al.  1998; Brinchann \& Ellis 2000; Papovich et al. 2001;
Shapley et al. 2001; D03; F03a) and
will be only briefly summarized here.  The main difference with
respect to the MA method is that the galaxy ages are also allowed to
vary at any $z$, with the only constraint that they are smaller than
the Hubble time at that redshift.

The set of template stellar populations adopted in the present work
are listed in Tab.~\ref{tabBC}. For each galaxy, the best--fitting SED
to our multiband $UBVRIzJK_s$ photometry is extracted from the grid at
the corresponding redshift with a $\chi^2$ minimization and used to
compute at once the stellar mass and all the rest frame
luminosities. In the following we refer to this as the ``Best Fit''
(BF) method.  A technique can be applied to estimate the confidence
levels on the estimated mass, taking into account the degeneracies
among the input parameters, as described in F03 and
in Appendix A.4.  We explicitly note that while the ``MA'' models are
designed to match the observed $K_s$ magnitude, the ``BF'' mass
estimates are derived from the model that indeed ``best fits'' the
global multicolor SED. In any case the average of the differences
between the observed $K_s$ magnitudes and those of the best fit model
is very small ($<0.01$~mag), with 10\% r.m.s. fluctuations.

   \begin{table}
      \caption[]{Parameters Used for the Library of Template SEDs}
         \label{tabBC}
     $$ 
         \begin{array}{p{0.3\linewidth}l}
            \hline
            \noalign{\smallskip}
            IMF & Salpeter\\
            \noalign{\smallskip}
            \hline
            \noalign{\smallskip}
            SFR $\tau$ (Gyr) & 0.1,0.3,0.6,1,1.5,2,3,4,5,7,9,13,15\\
            \noalign{\smallskip}
            \hline
            \noalign{\smallskip}
            log(Age) (yr) & 7,7.02,7.04...10.2^{\mathrm{b}}\\
            \noalign{\smallskip}
            \hline
            \noalign{\smallskip}
            Metallicities & 0.02Z_{\odot}, 0.2Z_{\odot}, Z_{\odot}, 2.5Z_{\odot}^{\mathrm{c}} \\
            \noalign{\smallskip}
            \hline
            \noalign{\smallskip}
            $E_{\rm B-V}$ & 0,0.03,0.06,0.1,0.15,0.2,...,1.0\\
            \noalign{\smallskip}
            \hline
            \noalign{\smallskip}
            Extinction Law & SMC  \\
            \noalign{\smallskip}
            \hline
          \end{array}
     $$ 
\begin{list}{}{}

\item[$^{\mathrm{a}}$] At each $z$, galaxies are forced to have ages lower than the Hubble time at that $z$.
\item[$^{\mathrm{b}}$] Models with metallicity=$0.02Z_{\odot}$ have been limited 
to log(Age)$\leq 9$; models with metallicity=$2.5Z_{\odot}$ have been limited 
to log(Age)$\geq 9$. 
\end{list}
   \end{table}

We are aware that, despite the wide parameter space covered by the
``BF'' grid of templates, there is no guarantee that the best-fit
solutions are, at least statistically, correct. First, several
simplified assumptions are used in building the library grid: the most
important are a monotonic, exponentially declining star-formation
history and a single metallicity for each SED template.  The adoption
of a universal IMF and a single extinction law (and in particular our
choice of the Salpeter IMF and the SMC law) may not be the most
appropriate. Moreover, some models within the grid may be not physical
(e.g. those implying large dust extinctions in absence of a
significant star-formation rate).

In order to remove the most obvious unplausible - or at least less
likely - models, we have not included in the libraries heavily
extincted models ($E(B-V)>0.2$) with no ongoing star formation
activity.

\subsubsection{The z=0 check}

Both MA and BF masses were also derived from a set of over 6000
galaxies using the publicly available multicolor catalogs of the SDSS
and 2MASS surveys, along with the SDSS redshifts, repeating a
procedure already adopted by Bell et al. 2003 to derive the local
stellar mass function.  This allows us to compare in a
self--consistent fashion the K20 stellar mass function obtained at
high $z$ with the BF method with the local one. To reduce the effects
of degeneracies among the models with short star--formation
time-scales, models with star--formation time-scales $\tau\leq 0.6$
and $z_{\rm form}<1$ were removed from the set of templates. Such a
choice matches also the model grid adopted by Kauffmann et al. 2003 in
their analysis of the SDSS spectra.  Such a selection in the grid
models, for self-consistency, has been applied also on the estimates
of the K20 sample, but without appreciable effects, since most early
type galaxies are at $z>0.5$ and their best fit ages are independent
of this selection.

\subsubsection{The case of dusty EROs}

We note that a small fraction of the K20 sample is made of the
so-called ``dusty'' ERO population (Cimatti et al. 2002c): these
objects are characterized by strong emission line features and require
large extinction (corresponding to $E(B-V)=0.6-1.1$ for a Calzetti et
al. 2000 law).  These objects are difficult to be modeled, since we
expect them to be made of different stellar populations with a complex
absorption geometry.  On the one side, the adoption of a Calzetti
attenuation curve is a coarse approximation, since is has been derived
only on local starbursts, with a maximal extinction ($E(B-V)=0.7$)
lower than the one inferred in our EROs, and it has been show to be
inapplicable to local Ultra Luminous IR starburst (Goldader et al
2000), although these objects are probably more extreme than our
EROs. On the other side, theoretical models allowing for a patchy,
complex dust geometry (Granato et al. 2000) suggest that the outcoming
attenuation curve may still be close to a single curve, similar to
Calzetti one.

For simplicity, we have used for these objects the same set of
templates as in Table 1, searching for the best fit only among the
star-forming models with $age/\tau \leq 2$, but adopting both
the SMC and the Calzetti curve, and performed two simple tests.
First, we verified that the estimated stellar mass does not depend
dramatically on the assumed extinction law, since the SMC--based
estimates are typically 20-30 \% higher than the Calzetti--based
estimates.  Second, we have found that the estimated star--formation
rates obtained adopting a Calzetti law are consistent with the X--ray
and radio luminosities of these objects (Daddi et al 2004b),
suggesting that we are not missing a significant fraction of their
star--formation activity. In the following, we shall therefore adopt
the masses estimated with the Calzetti law.

\subsubsection{The effects of different dust extinction curves}

The impact of different extinction curves on the mass estimates has
already been investigated by Papovich et al. (2001), D03 and F03, and
found to be small. In the HDFS, in particular, F03 found that adopting
the Calzetti extinction curve leads to mass estimates $\simeq 20$\%
lower than those estimated by adopting the SMC law, with a typically
worse $\chi^2$ at $z\leq2$. 
We have repeated the same exercise on the
K20 data set, finding again that the typical $\chi^2$ with a Calzetti
extinction curve is worse than that obtained with the SMC one, but
that the mass estimates are nevertheless similar, with an average
shift of 0.04 {\it dex} in the stellar mass (the SMC-based masses
being still larger than the Calzetti ones) with 0.2 {\it dex} of
dispersion. Likely, the difference between the two extinction curves
is lower in the K20 data set since it is richer in red, evolved
galaxies whose best fit spectra do not require a large amount
of extinction.  We have also verified that the stellar mass densities
are not changed significantly by changing the extinction curves.

On the basis of these considerations we believe that the use of the
SMC law is well justified, considering that our $K$-selected sample is
not expected to contain a large fraction of starbursts, with the
possible exception of the star-forming EROs which anyway have been
also fitted with a Calzetti law.

\subsubsection{The typical error on mass estimates}

We find that the typical error on the estimated masses is of the order
of $_{-40\%}^{+60\%}$ (Appendix A.4), in agreement with the similar
results in the HDFN (D03) and HDFS (\cite{fontana2003}) at the same
redshifts. Therefore, there appears to be a core level of degeneracy
in the input models that cannot be eliminated within this approach.
Nevertheless, this uncertainty is smaller than that affecting the
stellar mass estimates at $z\simeq 3$ (\cite{Papovich2001},
\cite{shapley}), since at $z\simlt 2$ we can rely on at least a
partial sampling of the rest-frame, near-IR side of the spectrum.

It is beyond the aim of the present work to describe in detail the
results of the fitting procedures for all the parameters involved.
Nevertheless, it is worth mentioning that we have checked that their
distributions appear to be astrophysically reasonable, such that we do
not expect large systematic biases in the mass estimates. For
instance, the resulting distribution of metallicity is peaked at the
solar value, with only 10\% of the objects at $Z=2.5Z_{\odot}$ and
about 25\% of them at $Z=0.2 Z_{\odot}$.  The average dust extinction
on the whole sample is $<\!E(B-V)\!>\simeq 0.2$, and
$<\!E(B-V)\!>\simeq 0.1$ for the objects spectroscopically classified
as early type.  The median (average) $z_{\rm form}$ is about 2 (2.8)
for spectroscopic early type and 1.3 (1.8) for spectroscopically late
type objects.  We have also found that the ``BF'' estimates reproduce the
amplitude of the 4000 \AA~ spectral break, as measured in our
spectroscopic sample (see Appendix A.3).

However, from simulations fully described in Appendix A.2, we find
that the derived galaxy ages can significantly underestimate the
actual ages in cases of complex star-formation histories, although
this does not affect the mass estimate by more than 25\%.  In particular,
in simulations with a secondary starburst on top of an exponentially
declining star formation, the mass is on average underestimated by
$\sim 25\%$ if the object is observed just during the starburst
itself.  But 1-2 Gyrs after the starburst the fits based on single
exponential laws are able to recover essentially all the stellar mass,
although the derived age may still be underestimated. This is
particularly important for the study of early type galaxies, that make
the bulk of the massive objects at $z\simeq 1$.

\subsection{Maximal Age vs. Best Fit Masses}

As expected, the MA models provide mass estimates that on average are
larger by a factor $\sim 2$ at low masses and $\sim 1.6$ at high
masses ($M_*\gsim 4 \times 10^{10}M_\odot$), although a large scatter
exists (see Appendix A.1). Such a mass difference is strongly
correlated with the age difference between the MA and BF estimates
(the median $z_{\rm form}$ for the total sample 
resulting from the ``BF'' approach is smaller than 2,
to be compared with the adopted fixed value $z_{\rm form}=20$ of the
MA models), while other parameters have a minor impact.  This assumed
high redshift for the start of star formation is
indeed the main difference between the ``Maximal Age'' MA models
and the BF models.

In addition, we have taken advantage of the
relation between MA and BF masses to obtain a ``BF'' mass estimate of
the 3 objects in the CDFS field (that are at $z=0.366, 1.087, 1.277$)
for which we do not have a full multicolor coverage, but only a MA
mass estimate from available $R-K_s$ color. 

Apart from systematic biases that would affect both methods in the same way
(see Section 5.5), and based also on the results of the simulations and of the
other tests described in Appendix A, we believe that considering the
results of both methods gives an idea of the existing uncertainties
and we will therefore consider both estimates in our subsequent
analysis.

\section{Galaxy stellar masses and \ml \ up to $z\simeq 2$}

\subsection{Galaxy stellar masses}

\begin{figure}[ht]
\resizebox{\hsize}{!}{\includegraphics{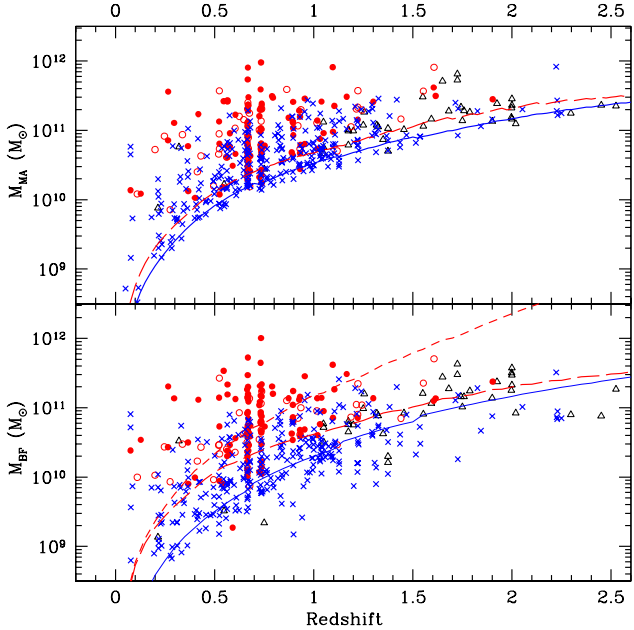}}
\caption{ Galaxy stellar masses in the K20 sample as a function of
redshift. Upper panel: estimates based on Maximal Age model. Lower
panel: estimates based on the Best Fit method. Filled circles: early
type galaxies; empty circles: early$+$emission type; crosses:
``star--forming'' type; triangles: objects with photometric redshifts.
A few very low redshift objects at $M_*<7\times 10^{8}M_\odot$ are
omitted. The short and long dashed lines correspond to the
completeness threshold, define by computing a maximally massive model
with $K_s=20$, considering dust and dust-free models, respectively
(see text for details).  Solid lines correspond to the selection
curves that we have applied when building the Galaxy Stellar Mass
Function, after application of the appropriate correction for
incompleteness as described in Sect. 5 and Appendix B.  }
\label{mass_z}
\end{figure}

The derived masses for the K20 sample are shown in Fig.~\ref{mass_z}
as a function of redshift. For each object, both the MA (upper panel)
and the BF estimates (lower panel) are shown.  It is immediately
apparent that very massive galaxies, in the range
$10^{11}M_{\odot}<M_*<10^{12}M_{\odot}$, are detected all the way to
$z\simeq 2$.  Besides the $M_* \simeq 10^{12} M_\odot$ galaxies within
the two $z\simeq 0.7$ peaks, the most massive galaxy with
spectroscopic redshift in the K20 sample is CDF1-633 at $z_{\rm
spe}=1.096$, $M_*=(4-8) \times 10^{11} M_\odot$, according to the BF
and the MA method, respectively.

In the HDFN, the upper envelope of the mass distribution appears to
decrease at high redshift (D03).
Such a trend is much less clear in the similar analysis of the HDFS
(F03), and even less so in the present K20 sample.

Conversely, the lower envelope is a result of the $K<20$ selection
criteria, that prevents less massive objects to be detected.  However,
as already discussed by D03 and F03, {\it IR--selected samples do not
strictly correspond to mass--selected ones}.  Indeed, at any Hubble time
(i.e. redshift), for each  $K$-band luminosity there is a range
in the possible $M_*/L$ ratios that is set by the range of allowable ages,
metallicities and dust extinctions of the observed stellar population.
This implies that at the low mass side the sample is progressively biased
against the detection of high \ml, such as old, passively evolving or
highly extincted galaxies.

Because of the uncertainties in the modeling of such objects, it is
difficult to define a clear mass threshold as a function of redshift.
A very conservative way to estimate it corresponds to the minimum mass
that a $K_s=20$ galaxy at any given redshift may have, within the
adopted set of spectral templates.  Such a threshold strongly depends
on the adopted library, and in particular on the allowed maximum
extinction.  For both BF library, this threshold is shown as short
dashed lines in Fig.~\ref{mass_z}. In this case, this threshold
corresponds to the maximum extinction allowed in the set of templates
(which is entirely arbitrary) and it would eliminate a large fraction
of the sample from the statistical analysis.  A more realistic
approach (already adopted by D03 and F03) is to consider only
dust-free, passively evolving models, such that the derived threshold
corresponds to the selection for early type galaxies.  In the case of
both the MA and BF libraries, this is shown as a long dashed line in
Fig.~\ref{mass_z}.  Our sample is therefore definitely incomplete
below this curve, where several objects of lower \ml \ (typically
star--forming galaxies) are still detected, and reasonably complete
above, except for strongly obscured sources.  The selection curves in
Fig.~\ref{mass_z}. shows that our sample is ``mass--complete'', except
for strongly obscured sources, down to objects of $M_*\simeq 3 \times
10^{10}M_\odot$ at $z\simeq 1$, and of $M_*\simeq 2\times
10^{11}M_\odot$ up $z\simeq 2$.

In principle, only objects above the long dashed lines of
Fig.~\ref{mass_z} should be used in statistical analyses that require
mass--selected samples, such as average \ml, mass densities or mass
functions. In the Section 5.1 and in Appendix B we will describe how
we have introduced a correction for the incompleteness in order to
extend the construction of the mass function to lower masses. Thanks
to this approach, we will be able to recover a significant fraction of
our sample: the corresponding selection curves are shown as solid
lines in Fig.~\ref{mass_z}.

\subsection{The \ml ratio}
\begin{figure}[ht]
\resizebox{\hsize}{!}{\includegraphics{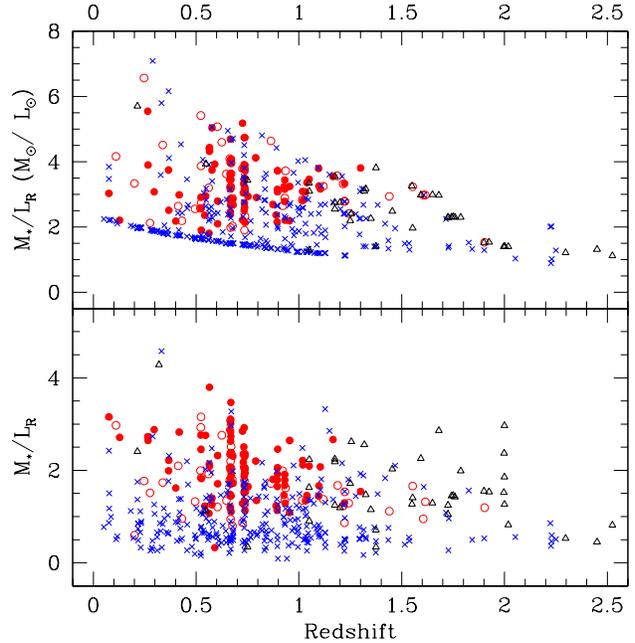}}
\caption{ $M_*/L_R$ ratios as a function of redshift, in the K20
sample. Upper panel shows the Maximal Age estimates, while lower one
shows the Best Fit ones. In both panels the symbols correspond to
different spectral types, as in Fig.~\ref{mass_z}.  }
\label{M_L_obs}
\end{figure}

Fig.~\ref{M_L_obs} shows the values of the $M_*/L_R$ ratio for the K20
sample of galaxies as a function of redshift, as derived from the MA
and BF procedures. Here, $L_R$ is the absolute luminosity in the
rest-frame $R$ band, in solar units, which is extracted from the
spectral template that best-fits the corresponding observables, and
that is well sampled by our multicolor photometry up to $z\simeq 2$.
For the reasons described above, the $M_*/L_R$ ratio tends to be
higher in the MA models than in the BF ones. Using the ``mass
complete'' sample described in Appendix B, we have obtained the
average $M_*/L_R$ ratio for the whole sample and for the two spectral
types as a function of redshift (in the redshift bins adopted to
compute the galaxy stellar mass function). The corresponding values
are shown in Table~\ref{mltab}, along with the corresponding
dispersions in the data.

In the redshift range including most of the galaxies in the sample
($0.2\leq z \leq 1.5$), a correlation exists between the
average \mlr \ ratio and the observed spectral type, with the
spectroscopic early type galaxies having on average a higher \mlr \
ratio with respect to the late type, star-forming ones. In
addition, we also detect an overall trend of decreasing \mlr \ with
increasing redshift, for both the early and the late type
galaxies.

\begin{figure}[ht]
\resizebox{\hsize}{!}{\includegraphics{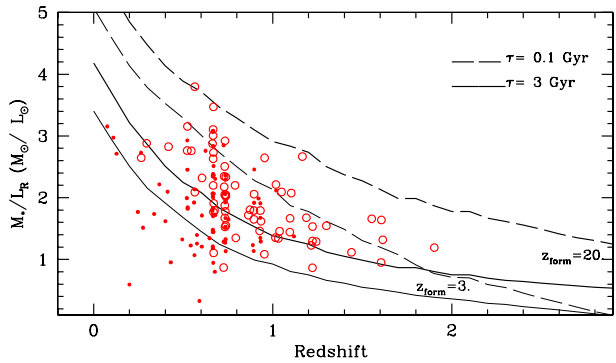}}
\caption{$M_*/L_R$ ratios as a function of redshift for
spectroscopically early and early$+$emission spectral type galaxies of
the K20 sample. Large hollow circles are for bright ($M_R<-22$) objects,
small filled circles for fainter ($M_R>-22$) ones.
Lines show the \mlr \
values computed with a set of single--exponential models with $z_{\rm
form}=3$ (thin lines) or $z_{\rm form}=20$ (thick lines) and
star--formation time-scales ranging of $\tau = 0.1$ Gyr (dashed lines) and
$\tau = 3$ Gyr (solid lines), all drawn with a Salpeter IMF and no dust
extinction.}
\label{MLave}
\end{figure}

It is of some interest to compare the observed \mlr \ ratio of early
type galaxies with that expected in simple cases of PLE models.  We
plot in Fig.~\ref{MLave} the \mlr \ for spectroscopically early and
early$+$emission types as a function of redshift, differentiating
between brighter ($M_R<-22$) and fainter ($M_R\geq-22$) objects, and
compare the observed evolution with a set of single-exponential models
with $z_{\rm form}=3$ and $20$ and $\tau = 0.1$ and $3$~Gyrs, all
computed with a Salpeter IMF and no dust extinction.  For simplicity, we
plot only the BF estimates. It is shown that the observed \mlr \
values are distributed over a significant range, suggesting that
coeval, single--exponential models are probably an oversimplified way
to describe the properties of spectroscopically early type galaxies.
A similar result was also presented on the EROs subsample (Cimatti et
al 2003), and is now extended to the whole early type population.

More interestingly, we show that the typical \mlr \ of brighter
objects is significantly larger than that of the fainter ones at low
and intermediate redshifts. This implies that, while the \mlr \ of the
bright population is consistent with either very short star--formation
time-scales or high $z_{form}\geq 3$, and some objects appear to
require both, the fainter population has experienced a more recent
history of assembly, witnessed by the larger $\tau$ and lower
$z_{form}$ required to reproduce the typical \mlr.  So, the most
luminous and massive galaxies appear to reach near completion first,
while less massive ones keep growing in mass till later times. This
{\it down-sizing} effect was first noted by Cowie et al. (1996) for
$z\simgt 1$, and then explicitly quantified by Brinchmann and Ellis
2000 and F03 (see also Kodama et al. 2004). This tendency continues
all the way to $z\sim 0$, where the star-formation rate per unit mass
{\it anti-correlates} with galaxy mass (Gavazzi et al. 1996; Kauffman
et al. 2003).

\begin{table} 
\caption[]{Average $M_*/L_R$ ratios as a function of redshift and
spectral type, computed in the mass complete--sample, as defined by
the thick solid curve of Fig.~\ref{mass_z}. {\it All} refers to all
galaxies in the K20 sample, including also those with photometric redshift
only.  {\it Early} refer to early and early+emission
spectroscopic types and {\it late} to late spectroscopic types, as
defined in Section 2. Only objects with spectroscopic redshift are
included in the last two columns. \ml~ ratio in the V and K bands are available
in electronic form at the web site {\sf http://www.arcetri.astro.it/$^{\sim}$k20}}
\label{mltab}
     $$ 
         \begin{array}{|l|ccc|ccc|ccc|}
            \hline
~  & \multicolumn{9}{c|}{\rm Maximal Age} \\

            \hline

{\rm Redshift}  &  \multicolumn{3}{c|} {\rm All} &  \multicolumn{3}{c|} {\rm Early} &  \multicolumn{3}{c|} {\rm Late} \\

            \hline
~ & \it n & \frac{M}{L_R} & \sigma & \it n & \frac{M}{L_R} & \sigma & \it n & \frac{M}{L_R} & \sigma \\
z < 0.2              & \it     9 &  2.83 &  0.82 & \it   3 &  3.13 &  0.98 & \it   6 &  2.67 &  0.78 \\
0.2 \leq z \leq 0.7  & \it   148 &  3.02 &  1.30 & \it  69 &  3.46 &  0.90 & \it  76 &  2.49 &  1.22 \\
0.7 \leq z \leq 1.0  & \it    89 &  2.93 &  0.91 & \it  45 &  3.31 &  0.73 & \it  44 &  2.54 &  0.93 \\
1.0 < z \leq 1.5     & \it    71 &  2.65 &  0.75 & \it  18 &  3.19 &  0.38 & \it  43 &  2.34 &  0.76 \\
 1.5 < z \leq 2.0    & \it    19 &  2.39 &  0.66 & \it   5 &  2.74 &  0.69 & \it   3 &  1.73 &  0.45 \\
            \hline

~ & \multicolumn{9}{c|}{\rm Best Fit} \\

            \hline

{\rm Redshift}  &  \multicolumn{3}{c|} {\rm All} &  \multicolumn{3}{c|} {\rm Early} &  \multicolumn{3}{c|} {\rm Late} \\

            \hline
~ & \it n & \frac{M}{L_R} & \sigma & \it n & \frac{M}{L_R} & \sigma & \it n & \frac{M}{L_R} & \sigma \\
z < 0.2              & \it    10 &  1.77 &  0.94 & \it   3 &  2.95 &  0.22 & \it   7 &  1.27 &  0.56 \\
0.2 \leq z \leq 0.7  & \it   104 &  1.92 &  0.83 & \it  61 &  2.18 &  0.66 & \it  42 &  1.48 &  0.82 \\
0.7 \leq z \leq 1.0  & \it    65 &  1.73 &  0.49 & \it  47 &  1.86 &  0.45 & \it  18 &  1.40 &  0.45 \\
1.0 < z \leq 1.5     & \it    38 &  1.78 &  0.54 & \it  16 &  1.61 &  0.45 & \it  16 &  1.81 &  0.62 \\
 1.5 < z \leq 2.0    & \it    15 &  1.53 &  0.53 & \it   5 &  1.35 &  0.30 & \it   2 &  1.26 &  0.43 \\
            \hline
            \hline
          \end{array}
     $$ 
   \end{table}

\section{Galaxy Stellar Mass Functions}

\subsection{The construction of the Galaxy Stellar Mass Functions }

Once the stellar mass has been obtained for each galaxy in the
sample, the building of the corresponding Galaxy Stellar Mass Function
(GSMF) follows the traditional techniques used for
luminosity functions.  We apply here both the classical $1/V_{\rm max}$
formalism to obtain binned distributions, and the Maximum Likelihood
technique to estimate the best-fit Schechter parameters, using the
recipes described in Poli et al. (2001, 2003) and Pozzetti et al. (2003), and
references therein.

With respect to the standard procedures adopted for luminosity
functions, an improvement is necessary to account for the
incompleteness in the {\it mass} census of galaxies. The adopted
procedure to obtain the correction fractions for incompleteness is
fully described in Appendix B. This correction factor is then
applied to the volume element $V_{\rm max}$ of any galaxy, both in the
$1/V_{\rm max}$ binned GSMF as well as in the Schechter best-fits.

Another aspect requiring particular attention is the comparison of GSMFs 
derived at the various redshifts with
the local, $z\sim 0$ GSMF. In the following, we shall make use of the local
GSMFs  derived by \cite{cole2001} and
\cite{bell} with the same Salpeter IMF.  The former authors estimate the
spectral type by the observed $J-K$ color, while the latter ones use the
complete set of $ugrizJHK$ SDSS+2MASS colors. Since both authors use
a set of spectral synthesis models with star-formation rate 
peaking at high redshifts, the two GSMFs agree quite well with each
other, especially for $M_* > 10^{10} M_\odot$, and the comparison with
our MA estimates at higher redshift is fair, since our MA method
mimics exactly the \cite{cole2001} models.

Conversely, the same may not hold for our BF estimates, that have no
constraint on age and typically yield lower $z_{\rm form}$ for the K20
objects.  To estimate the effect, we have repeated the \cite{bell}
procedure, building a sample of 6332 SDSS+2MASS local galaxies, for
which we obtained both the MA and BF stellar masses with our
recipes. For this local sample we find that the BF mass estimates are
on average lower than MA by only about 20\%, and even less for more
massive objects. Full details of this analysis are described in
Appendix A.5.  Using this result to statistically convert the
\cite{cole2001} mass function to a BF one, we have obtained a
``BF--scaled'' local GSMF that only marginally departs from the
original of \cite{cole2001} in the massive tail, and that we will use
to compare with our ``BF'' results at higher $z$.

Finally, the intrinsic uncertainty in the estimate of the stellar mass
must be taken into account when computing the error budget in the
GSMF. At this purpose, we have performed a set of MonteCarlo
simulations where the input mass catalogs were randomly perturbed,
allowing each galaxy in the sample to move around its best fit values
of a quantity specified (for each galaxy) by the error analysis
described in Appendix A.4. This includes also the uncertainty due to
the redshift for galaxies with photometric redshifts only. We
simulated 200 such catalogs, and computed the resulting GSMF. The
dispersion in the derived values (both for the binned valued as well
as for the fitted Schechter parameters) has been added in quadrature
to the standard Poisson noise for each GSMF. We remark that a
more global uncertainty - not shown in the following figures - is
related to cosmic variance, that is much more difficult to treat.  As
we discuss in better detail in Sect. 5.5, the error budget due to
cosmic variance is around 20-40\% (depending on the assumed galaxy
correlation length) of the total number densities.

\subsection{The evolution of the Galaxy Stellar Mass Functions}

\begin{figure*}
\centering
\resizebox{\hsize}{!}{\includegraphics[angle=-90]{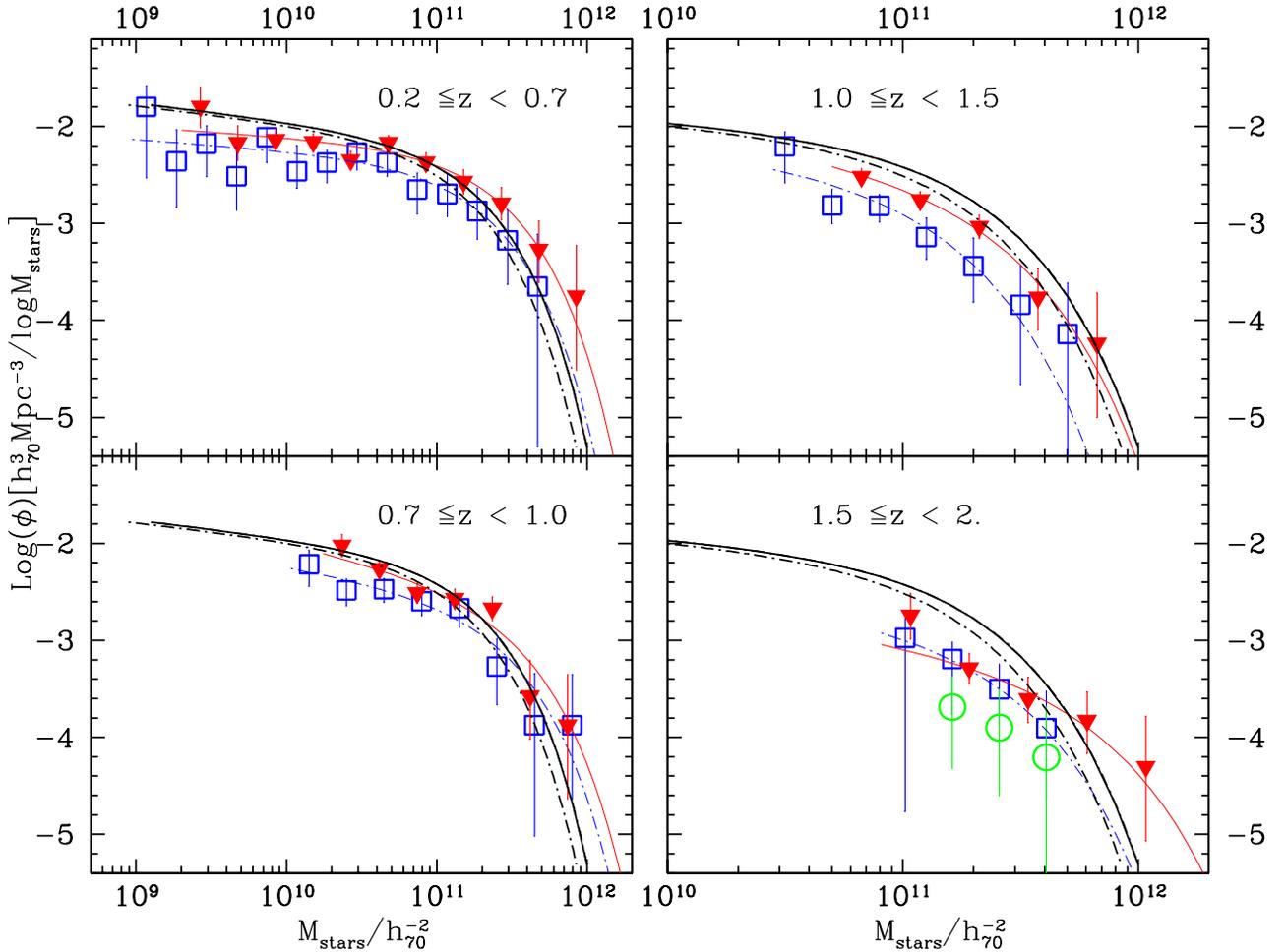}}
\caption{ Galaxy stellar mass functions in the K20 sample, in four
redshift ranges. Different symbols correspond to different methods
adopted to estimate the stellar mass: triangles represent the Maximal
Age estimates, and squares the Best Fit estimates. In the highest
redshift bin, large hollow circles correspond to the GSMF measured by
using only objects with spectroscopic redshift.  Thick lines
correspond to local galaxy mass functions: the solid line is
the local galaxy mass function by \cite{cole2001}, the dot--dashed one is
the same GSMF renormalized to our ``BF'' method (See text and
appendix A for details).  The same line types have been used to
represent the Schechter fits to our ``extended'' galaxy mass
functions (thin lines).}
\label{massfun}
\end{figure*}

The resulting stellar mass functions of the K20 sample are shown in
Fig.~\ref{massfun}. We have divided the sample into four redshift
bins: 0.2--0.7; 0.7--1.0; 1.0--1.5; and 1.5--2.0, having chosen the
two lowest redshift bins in such a way to distribute in two different
bins the impact of the two redshift peaks at $z=0.67$ and $z=0.73$,
and to have a similar comoving volume (about $2.5\times
10^4$Mpc$^3$). The two high redshift bins have also a comparable
comoving volume ($7\times 10^4$Mpc$^3$ and $8.2\times 10^4$Mpc$^3$,
respectively). We have restricted the estimate of the GSMF at $z<2$,
given the small number of objects and the limited mass range covered
by our sample beyond $z\simeq 2$.  The average redshift of the
galaxies in each bin is $0.52,0.83,1.17$ and $1.72$, respectively.  In
terms of cosmic time, for the adopted cosmology these redshift
intervals correspond to an age of the Universe of $11.3-7.5$,
$7.5-6.1$, $6.1-4.6$ and $4.6-3.6$, respectively. For each redshift
bin we have constructed the GSMF following the two different methods
described in the previous sections: Maximal Age and Best Fit. For each
bin, error bars include both the Poisson noise (computed using the
exact formulas for low counts given in Gehrels (1986)) as well as the
uncertainty in mass estimates, obtained with the Monte Carlo procedure
described above.

In the highest $z$ bin, the contribution by objects which have only a
photometric redshift is significant. In order to provide a strict
lower limit to the total GSMF, in this redshift bin we have also used
only the spectroscopic sample with no correction for incompleteness
(large hollow circles in Fig.~\ref{massfun}).

\begin{table}
\caption[]{Best fit Schechter parameters of the Galaxy Stellar Mass
Functions in the K20 survey. Errors indicate $1\sigma$ confidence
levels. Parameters without error have been fixed to the best fit value
of the lower $z$ bin.}
\label{alpha}
     $$ 
         \begin{array}{|l|l|ccc|}
            \hline
 \multicolumn{5}{|c|}{\rm  Schechter~Parameters~for~the~GSMF} \\

            \hline

{\rm Redshift}  & ~ & \alpha & M^* & \phi^* \\

            \hline

0.2 \leq z < 0.7  & {\it BF} & -1.11\pm0.10  & 11.22^{+0.13}_{-0.12}  & 0.00182  \\
0.2 \leq z < 0.7  & {\it MA} & -1.10 \pm0.10  & 11.32 \pm0.11  & 0.00252  \\
            \hline

0.7 \leq z < 1.0  & {\it BF} & -1.27\pm0.23  & 11.37^{+0.22}_{-0.21}  & 0.00110  \\
0.7 \leq z < 1.0  & {\it MA} & -1.36 \pm 0.23  & 11.44 \pm0.19  & 0.00132\\

            \hline

1.0 \leq z < 1.5  & {\it BF} & -1.27  & 10.99 ^{+0.16}_{-0.1}  & 0.00147\\
1.0 \leq z < 1.5  & {\it MA} & -1.36  & 11.20 \pm0.08  & 0.00150 \\

            \hline

 1.5 \leq z < 2.   & {\it BF} & -1.27 & 11.24 ^{+0.38}_{-0.18}  & 0.00067  \\
 1.5 \leq z < 2.   & {\it MA} & -1.36 & 11.63 \pm0.17  & 0.00026  \\

            \hline
            \hline
          \end{array}
     $$ 
   \end{table}

At each redshift, the resulting GSMFs are compared to the local one by
\cite{cole2001} for the same Salpeter IMF, as well as to the one
``rescaled'' to our BF estimates (see appendix A5,
Fig.~\ref{localM}). For simplicity, we do not plot the \cite{bell}
local GSMF, that is in excellent agreement with \cite{cole2001} mass
function.

In each panel of Fig.~\ref{massfun} we also plot the Schechter fits of
the ``extended'' GSMFs. We remark that the Schechter parameters
derived by using the ``strictly complete'' or the ``extended''
selection criteria are statistically consistent with each other.
Unfortunately, at $z>1$, we cannot reliably estimate the slope of the
GSMF because of the small range in mass covered by our sample. For
this reason, we have assumed in these bins the same slope that we
observe at $0.7 \leq z < 1$, which may bias the estimate of the
characteristic mass in the high $z$ bins. For this reason, and given
the size of the resulting statistical errors, one should not read too
much in these values, that should be primarily used as description of
our dataset in the range that we actually observe.  The corresponding
best fit values are given in Table \ref{alpha}, with their $1 \sigma$
uncertainties.

The most relevant results that emerge from Fig.~\ref{massfun} are the
following:

First, we note that the overall agreement between the GSMF derived with the BF
and the MA procedures is fairly satisfactory.  Albeit MA
estimates provide higher normalizations than the BF estimates, as
expected by their typically larger masses, the resulting scenarios are
not significantly dependent on the adopted method of mass
determination, and so are the conclusions that we shall draw in the
following.

\begin{figure}[ht]
\resizebox{\hsize}{!}{\includegraphics{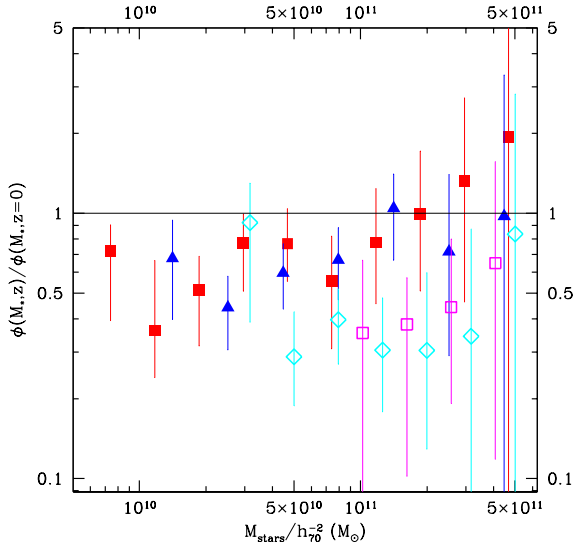}}
\caption{ Ratio between the observed Galaxy Stellar Mass Functions in
the K20 sample and the local GSMF. Data points refer to the ratio of
the ``Best Fit'' values to the corresponding ``rescaled'' local GSMF.
Filled symbols refer to points at $z<1.0$ (squares: $0.2<z<0.7$;
triangles $0.7<z<1.0$); open symbols to $z>1.0$ (diamonds:
$1.0<z<1.5$, squares: $1.5<z<2$).  The data points at $1.5<z<2$
include objects with photometric redshifts.}
\label{MF_rapp}
\end{figure}

Up to $z\simeq 1$ there appears to be only {\it a very mild evolution
of the GSMF} (see Fig.~\ref{massfun} and Fig.~\ref{MF_rapp}), as
suggested by the corresponding near-IR luminosity function (Pozzetti
et al. 2003).  Both the MA- and the BF-derived GSMFs are in agreement
with the local ones, at least in the range $10^{10}M_\odot < M_* < 2
\times 10^{11}M_\odot$, where our statistics is reasonably accurate.
The low-mass end of the GSMF is rather flat in the first redshift bin,
where it is well determined ($\alpha=-1.1 \pm 0.1$), quite consistent
with the local estimates ($\alpha=-1.18\pm0.03$) (Cole et al. 2001);
in addition, also the characteristic masses in the Schechter function
are consistent with local estimates.

At higher redshifts, $z>1$, there appears to begin a decrease in the
normalization of the GSMF, which is particularly remarkable if we take
into account the relatively small range of cosmic time resulting from
our sampling. The decrease is particularly evident for $M_*\simeq
10^{11}M_\odot$, and is approximately constant up to $z\simeq 2$.
This decrease in normalization is also shown in Fig.~\ref{MF_rapp},
where we plot the ratio between the GSMF at the various redshifts and
the local GSMF for the BF case.  The number density of objects around
$M_*\simeq 10^{11}M_\odot$ is (70-80\%) of the local value up to
$z\simeq 1$, while it decreases to (30-40\%) of the local value in the
two higher redshift bins.  This suggests that the mass assembly of
objects with mass close to the local characteristic mass was quite
significant between $z=2$ and $z=1$, and was essentially complete by
$z\simeq 1$.

The evidence presented in Sect. 4.2 of a differential evolution of
early type galaxies, with more luminous (i.e. more massive) galaxies
having formed earlier than less luminous one, should be reflected in a
flattening of the observed GSMF at high $z$. In our data there is a
tentative suggestion of this, since the massive tail of the GSMF
appear to evolve in a slower fashion than the fainter. Given the
present statistics, unfortunately, much wider surveys are required to
confirm this potentially important item.

\subsection{The GSMF for different spectral types}
\begin{figure}
\centering
\resizebox{\hsize}{!}{\includegraphics[angle=0]{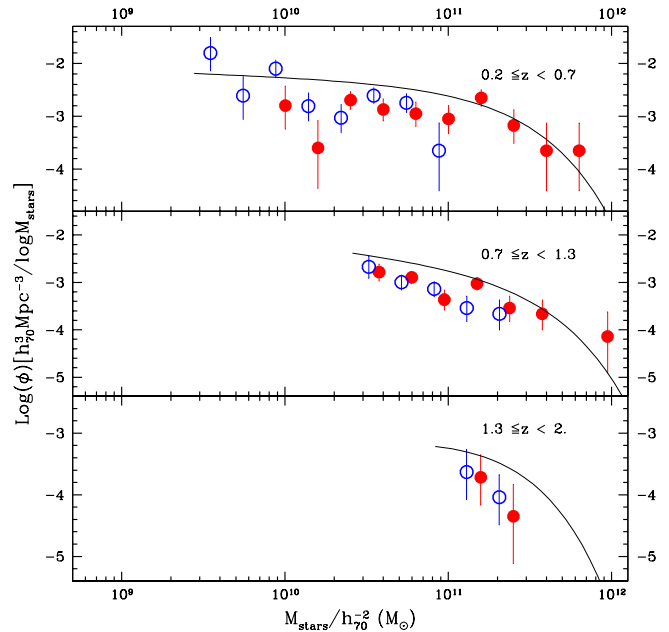}}
\caption{ Galaxy stellar mass functions in the K20 sample for
different spectral types. Empty points correspond to late
spectral type, filled to early spectral type.  The solid lines
show the Schechter fits to the {\it total} GSMF of our sample at the
corresponding redshifts.  }
\label{massfun_type}
\end{figure}

Fig.~\ref{massfun_type} shows the GSMF for the two main broad galaxy
spectral types, early and late (we remark that early$+$emission type
have been omitted). Since only objects with spectroscopic redshift are
included, these were grouped into three bins rather than four as in
Fig.~\ref{massfun}, thus ensuring a fairly good statistics in each
bin. We remind that the spectroscopic completeness of the K20 survey
is particularly high ($\simeq 92\%$), such that incompleteness effects
should play a minor role here. In any case, one expects galaxies
without a spectroscopic redshift to be preferentially at $z\simgt
1.3$, for which the most prominent spectral features have moved out of
the observed spectral range. For $z<1.3$, indeed, we estimate the
spectroscopic completeness to be 97\%, given the distribution of the
photometric redshifts of the unidentified objects.

We find that there is a clear difference in the GSMF for the two
spectral types (Fig.~\ref{massfun_type}).  Up to $z\simeq 1.3$, early
type galaxies dominate the mass distribution, and in practice provide
the whole contribution to its most massive side. This can be
qualitatively appreciated by observing the GSMF of
Fig.~\ref{massfun_type} and by observing that all the more massive
galaxies at these redshift bins are early type.  Although the latter
evidence may be somewhat affected by the prominent structures at
$z\simeq 0.7$, that are dominated by early type galaxies, a scrutiny
of our catalogs reveals that early type galaxies are the most massive
galaxies also outside the structures (see Fig.~\ref{mass_z}).
Quantitatively, we find indeed that the stellar mass density due to
massive galaxies (i.e. $M_*\geq 10^{11}M_\odot$) at $z\simeq 0.45$ and
$z\simeq 1$ is of 1.68 and 1.08~$10^{8}M_*$Mpc$^{-3}$, of which 85\% and
69\%, respectively, is due to early type galaxies, and 0\% and 16\% is
due to late type galaxies.
When integrated over the whole observed range early type galaxies provide
about 60\% of the whole stellar mass density at $z<1.3$.
A similar behavior was found in
the local GSMF by \cite{bell}, where the early-type/late-type
classification was based on morphology.

In the highest redshift bin, star--forming galaxies appear to
contribute a significant fraction of the massive tail of the GSMF.
The measured stellar mass density at $1.3<z<2$ due to star-forming
galaxies with $M_*\geq 10^{11}M_\odot$ is of the order of $1\times
10^{7}M_\odot/$Mpc$^{-3}$, corresponding to about 21\% of the total
mass density in the same mass range, an amount nearly identical to the
contribution of early type galaxies (22\%).  This figure is a {\it
lower limit} to the contribution of late-type galaxies to the total
stellar mass density, since it would correspond to assuming that all
the galaxies without spectroscopy in this redshift bin are passively
evolving objects (which implies that all photometric redshifts are
correct in this range and that all unidentified objects are
spectroscopically early type).

Irrespective of the nature of the unidentified objects, this result
suggests that a global physical change occurs at larger and larger
$z$, with an increasing fraction of the stellar mass density being
contained in actively star--forming objects.
This is also supported by the {\it upper} limit to the global stellar
mass density in passively evolving galaxies at $z\simeq 3$, that is
about 40\% of the total (at $z=\simeq 3$) in the HDFS (F03).

\subsection{The cosmological evolution of the  mass density}
\begin{figure}[ht]
\resizebox{\hsize}{!}{\includegraphics{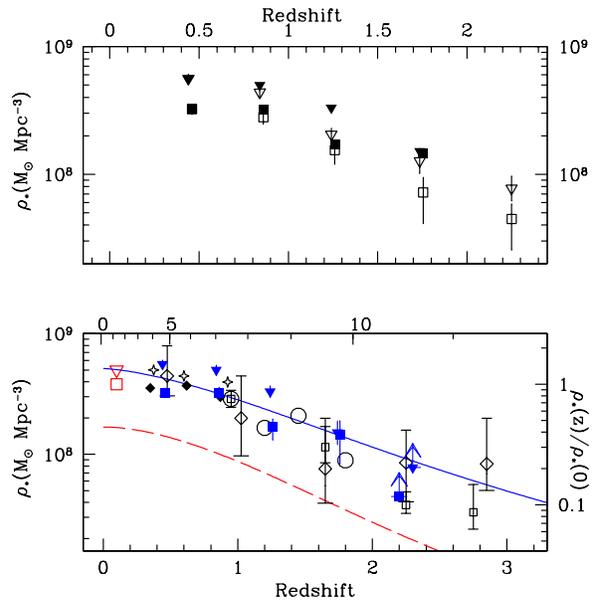}}
\caption{ Evolution of the cosmological mass density as a function of
redshift. {\it Upper panel}: Observed cosmological mass density as
observed in the K20 data. Squares correspond to BF estimates,
Triangles to MA estimates.  Empty points represent the {\it observed}
values, with the corresponding Poisson noise. Filled points represent
the values corrected for incompleteness, as obtained by integrating
the mass function over the whole range $10^{8} M_\odot \leq M_* \leq
10^{13} M_\odot$.  Error bar are computed from the uncertainties in
the best--fit parameters of the Schechter Function.  {\it Lower
panel}: The global evolution of the cosmological mass density from
$z=0$ to $z=3$ as observed from the K20 and other surveys.  Filled
points represent the total values from the K20 survey: squares are the
``Best Fit'' values and triangles the ``Maximal Age'' estimates. The
points at $<z>=2.25$ are lower limit since they are affected by
incompleteness. The large open triangle is the local value by Cole et
al. 2001, while the large open square is the same local value rescaled
to the BF technique. Open exagones are from Cohen et al. 2001, open
triangles from Brinchmann and Ellis 2000, open squares from D03,
diamonds from F03, empty circles from Glazebrook et al. 2004. Error
bars in the latter two surveys take into account for the systematic
uncertainties among the methods adopted, as in the case of our MA and
BF approaches. The two lines are the mass densities expected by
integrating the star formation histories of Steidel et al. 1999 for
two different extinction coefficients (dashed: $E(B-V)=0$; solid:
$E(B-V)=0.15)$. }
\label{massden}
\end{figure}

The upper panel of Fig.~\ref{massden} shows the stellar mass density
$\rho_*(z)$, as derived for the K20 sample with the two different
methods (BF and MA), both from the {\it observed} data, as well as
from the incompleteness-corrected GSMFs, i.e., integrating the
best--fit Schechter functions (Table \ref{alpha}) over the whole range
$10^{8} M_\odot \leq M_* \leq 10^{13} M_\odot$. The same quantites
are given in Tab.\ref{tabden}. The correction is
marginal at $z<1$, but becomes a factor of $\sim 2$ at $z\geq 1.5$,
where this exercise is obviously prone to statistical as well as
systematic errors due to the limited mass range of the observed GSMF.
Fig.~\ref{massden} gives the statistical errors due to the
uncertainties in the best-fit parameters of the Schechter function.

In the case of the two high redshift bins, where we have fixed the
slope of the GSMF (see Tab.\ref{alpha}), these error estimates do not
take into account the uncertainty in the slope of the GSMF: this
systematic effect is discussed and quantified in Sect. 5.5, as 
well as the global uncertainty due to cosmic variance.

The global evolution of the stellar mass density from the K20 sample
is reported again in the lower panel of Fig.~\ref{massden}, along with
the results from other surveys at $z\simeq 1$ (Brinchmann and Ellis
2000; Cohen 2002), as well as at higher $z$ from Glazebrook et
al. 2004 and from the HDFN (D03) and HDFS (F03).  The K20 survey data
show that the stellar mass density at $z\simeq 1.5-2$ is about 35\% of
the local value, with a lower limit (given by the strictly observed
quantity) of 20\%. This is well placed within the global trend that
witnesses a fast increase of the stellar mass density from
$20^{+20}_{-10}\%$ of the local value at $2<z<3$ to about unity at
$z\leq1$, i.e. in a relatively short span of cosmic time.  At $z>2$,
we cannot compute a reliable estimate of the total mass density since
most of the objects fall below the completeness threshold. For this
reason, we compute the same quantity on the observed data only, and
represent it as a lower limit.  Interestingly, this quantity seems to
support the higher values observed in the HDFS rather than those in
the HDFN.

While the current, direct estimate is still affected by sizable
uncertainties, it is worth considering the indirect estimate of the
fraction of the stellar mass already in place at high redshift which
is provided by the low redshift ``fossil'' evidence. Indeed, from the
fraction of the local stellar mass locked in passively evolving
spheroids ($\sim 50-75\%$; e.g., Persic \& Salucci 1992; Fukugita et
al. 1998; Benson et al. 2002), and from the high redshift of formation
($z\simgt 3$) for the bulk of stars in such spheroids (Renzini 1999a),
it has been inferred that at least $\sim 30\%$ of the local stellar
mass may well be already in place by $z=3$ (Renzini 1998, 2003).

\begin{table}
\caption[]{Observed and total stellar mass density in the K20 survey,
as a function of redshift. Reference value are given for the BF
estimates; lower errors  take into account the Poisson noise and the
uncertainties in the mass estimates. Upper limits have been computed
with respect to the values provided by the MA method, to provide an estimate
of the global uncertainties.}
\label{tabden}
     $$ 
         \begin{array}{|l|ccc|}
            \hline
 \multicolumn{4}{|c|}{\rm Stellar~Mass~Density~in~the~K20~survey } \\

            \hline
            \hline

{\rm Redshift}  & {\rm Observed^{\mathrm{a}}} & {\rm Total^{\mathrm{b}}} & M_* \simeq 10^{11} M_\odot^{\mathrm{c}}   \\
            \hline
~ & log(M_*/M_\odot) &log(M_*/M_\odot) &log(M_*/M_\odot) \\
            \hline

0.2 \leq z < 0.7  & 8.51 ^{+0.24}_{-0.04}  & 8.51 ^{+0.26}_{-0.04}  & 8.32^{+0.28}_{-0.04}   \\
            \hline

0.7 \leq z < 1.0  &  8.44^{+0.20}_{-0.05} & 8.50^{0.23}_{-0.05}    & 8.29^{+0.20}_{-0.04}  \\

            \hline

1.0 \leq z < 1.5  &  8.19^{+0.13}_{-0.11} & 8.23^{+0.33}_{-0.11}  & 7.88^{+0.39}_{-0.1}  \\

            \hline

 1.5 \leq z < 2.   & 7.86^{+.24}_{-.24} & 8.16^{+0.11}_{-0.24}  & 7.93^{+0.1}_{-0.22}  \\

            \hline

 2 \leq z < 2.5^{\mathrm{d}}   &  7.65^{+.24}_{-.24} & -  & -  \\

            \hline
          \end{array}
     $$ 
\begin{list}{}{}
\item[$^{\mathrm{a}}$] On the complete sample
\item[$^{\mathrm{b}}$] Computed extending the Schechter fits from  $ 10^{8} M_\odot$ to  $ 10^{} M_\odot$.
\item[$^{\mathrm{c}}$] Computed for objects with $5\times 10^{10} M_\odot \leq M_* \leq 5\times 10^{11} M_\odot$.
\item[$^{\mathrm{d}}$] This value is actually a lower limit: see text for detail.
\end{list}

   \end{table}

Following the approach of Pozzetti et al. 1998, D03 and Cole et
al. (2001), the lower panel of Fig.~\ref{massden} shows a comparison
of $\rho_*(z)$ from various surveys with the same quantity as obtained
by integrating over time the star-formation rate density from Steidel
et al. (1999), with two different assumptions for the dust extinctions
(i.e., $E(B-V)=0.0$ and 0.15).  There appears to be a good agreement
between the observed evolution of $\rho_*(z)$ and the integrated
star--formation history due {\it only} to UV bright galaxies, after
allowing for a reasonable amount of dust.

Given the uncertainties in the estimates of both the stellar masses
and the UV--corrected star--formation rates, it is still possible to
allocate space for other significant contributors to the global
star--formation rate, as could be the case of dust--enshrouded
sources. However, the overall agreement, and in particular the match
with the local value of the stellar mass density, as well as the
consistency with the luminosity densities at different wavelengths up
to $z\simeq 1$ (Madau, Pozzetti and Dickinson 1998), appear to support
the notion that the UV selection is able to recover a major fraction
of the star-formation activity at high redshift.

\subsection{Systematic uncertainties}

In closing this section it is worth emphasizing that several
systematic uncertainties still affect the mass estimates of individual
galaxies as well as the current estimates of the global mass density
at high redshift. Such uncertainties are cursorily listed below.

$\bullet$ {\bf Initial Mass Function}. The Salpeter IMF
($\phi(M)\propto M^{-2.35}$ from 0.1 to 100 $M_\odot$) was adopted to
allow a direct comparison with other observational or theoretical
GSMFs. However, all empirical determinations of the IMF indicate that
its slope flattens to $\sim -1.35$ below $\sim 0.5\; M_\odot$ (Kroupa
2001), including data both in the Galactic disk (Gould et al. 1996) and in
the Galactic bulge (Zoccali et al. 2000). Compared to the single-slope
Salpeter IMF, such a two-slope IMF would give masses about a factor of
2 smaller, for a given galaxy luminosity. However, this correction
applies by about the same factor at all redshifts explored in this paper,
and the relative evolution of the GSMF is not affected. Of more
concern is the possibility of a different slope for $M\simgt
M_\odot$. The use of a {\it top-heavy} IMF would appreciably reduce
the estimated masses of actively star-forming galaxies at high
redshifts, and would appreciably modify the evolution of the GSMF
derived from the present data.  On the other hand, a top-heavy IMF
would largely over produce metals compared to their observed amount in
galaxy clusters (Renzini 1999b), while probably leading to a too small
star mass density in the local universe.  Nevertheless, a more
systematic exploration of other assumptions on the IMF may have been
in order, but this goes beyond the scope of the present paper.

$\bullet$ {\bf Spectral coverage}. 
As the redshift increases, a bluer and bluer portion of the rest-frame
spectrum is used, which is progressively dominated by the young-age,
low-mass components of galaxies, rather than by the older, more
massive components. This might lead to an underestimate of the stellar
mass at high redshift. A quantitative estimate of this effect should
soon become possible as the Spitzer observatory will directly
provide rest-frame $K$-band luminosities all the way to $z\sim 3$ for
galaxies in the GOODS fields (Dickinson 2002).

$\bullet$ {\bf Highly obscured objects.} A $K$-band selection of
galaxies to trace the build-up of stellar mass is certainly the least
biased approach to the problem, yet one not totally free from a
selection bias. Highly obscured objects at high $z$, such as SCUBA
sources (Chapman et al. 2003), may well be fainter than our threshold
at $K_s=20$ and nonetheless they may contain a sizable fraction of the
stellar mass in the highest redshift bin, while experiencing extreme
starbursts. Again, this bias would introduce an underestimate of the total
stellar mass at high redshift.

$\bullet$ {\bf Shape of the Mass Function.} The incomplete coverage of
the mass function increases with redshift, hence the uncertainty in
the slope $\alpha$ of the GSMF increases with redshift (see Table 3),
such that we have been forced to fix the slope at $z>1$. In the case
that the high $z$ GSMF is steeper than our assumption, this would lead
to an underestimate of the stellar mass density in the two high $z$
bins, and to a (slight) overestimate in the opposite case. For
instance, the total mass density that we estimate in the redshift
range $z=1.0-1.5$ would result $\rho = 10^{8.21}$,$
10^{8.34}$,$10^{8.52}$ if the slope of GSMF is fixed to $\alpha =
-1.18,-1.5, -1.8$, respectively.  Only substantially deeper data could
improve the present estimates of the GSMF.

$\bullet$ {\bf Cosmic variance.} While some 10 times larger than HDF,
the area explored by the K20 survey is still quite small, likely to be
subject to sizable fluctuations in the number density of
highly-clustered massive galaxies. For example, field-to-field
variance is detected among COMBO-17 fields (Bell et al. 2004), each
of them almost 20 times the area covered by the K20 project.  Indeed, as
mentioned in Section 2, the K20 sample exhibits redshift peaks that
appear to be significantly above the average distribution at $z\leq
1$, and strong clustering among early type red galaxies (EROs) is
detected in our sample at $z\simeq 1$ (Daddi et al.  2002).

For the reasons described in Section 2, however, the number of these
structures is reasonably within the expectation, such that we do not
have any reason to remove them nor any statistical argument to
re-normalize their contribution.

Even in this case, the cosmic variance due to galaxy clustering
may affect our results. To estimate this effect, we have assumed two
possible values for the galaxy correlation lenghts, $r_0 = 5h
^{-1}$Mpc and $r_0 = 10 h ^{-1}$Mpc, respectively, and computed the
expected variance using Eq. 8 of Daddi et al. 2000. The value $r_0 =
5h ^{-1}$Mpc has been derived from an analysis of the clustering
observed in the K20 sample itself, of which we shall provide the
details elsewhere, while $r_0 = 10 h ^{-1}$Mpc is a safe upper limit
taken from the EROs clustering amplitude (Daddi et al. 2001).  We find
that the relative variances in our redshift bins are typically of
about 20-25\% in the former case, and 40\% in the latter. If we
restrict the computation to massive objects ($M_*\geq
10^{11}M_\odot$), which dominate the mass density, the expected
variance are only slightly larger, 30\% and 45\% respectively.

Ultimately, surveys over much wider areas will be necessary to fully
average out the impact of cosmic variance.  As a simple check,
however, we have analyzed independently the stellar mass densities and
GSMF in the CDFS or Q0055 field, that are separated by more that 45
degrees, finding a good agreement.  In our redshift bins, the scatter
between the stellar mass densities in the Q0055 and CDFS is typically
of 0.1 {\it dex}, and the resulting GSMFs are consistent. In
particular, the number density of massive galaxies and the decrease of
the GSMF at $z>1$ are observed in each individual field.  In addition,
we note that also the GSMFs recently obtained with the wider MUNICS
survey (Drory et al. 2004) result to be in excellent agreement with
our, in the range where they overlap.

\section{The comparison with $\Lambda$--CDM galaxy formation models}

The aim of this section is to undertake a comparison of the present
findings on the evolution of the galaxy population up to $z\sim 2$ with
the available predictions of some renditions of the CDM paradigm of
galaxy formation and evolution. This comparison will include both
semi-analytical as well as fully hydro dynamical models.

As mentioned in Section. 2, our total ``Kron'' magnitudes are prone to
systematic underestimates of the total flux, by an amount that depends
on the morphology, sampling, redshift and S/N of the objects.  In
order to coarsely correct for this ``photometric bias'', following
Pozzetti et al. (2003) we have shifted all the empirical GSMF by 20\%
up in mass before proceeding to a comparison with theoretical
models. Indeed, such models provide {\it total} galaxy luminosities,
unaffected e.g., by the systematic bias introduced by systematics in
the measured quantities.

\subsection{Overall consistency with the CDM paradigm}

We first examine whether the large number of massive galaxies that we
observe at high $z$ is consistent with the fundamental properties of
$\Lambda$--CDM hierarchical scenarios, where
\begin{figure}[ht]
\resizebox{\hsize}{!}{\includegraphics{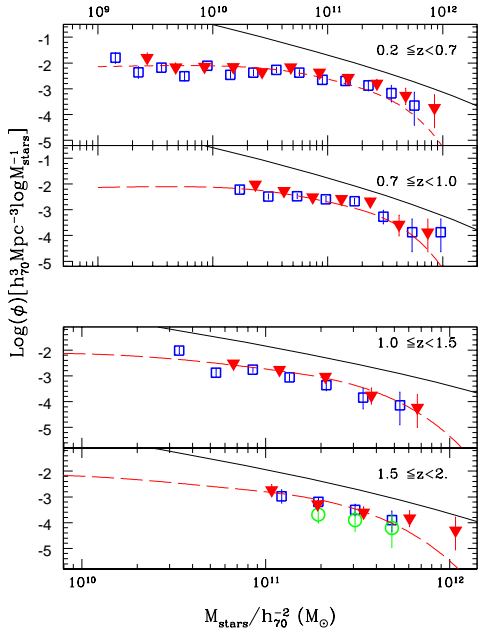}}
\caption{ Observed Galaxy Stellar Mass Functions in the K20 sample,
compared with the baryonic mass function in a $\Lambda$--CDM
hierarchical scenarios and with PLE predictions for the galaxy stellar
mass function.  The redshift ranges and symbols are the same as in
Fig.~\ref{massfun}.  In particular, hollow circles represent the
stellar mass function computed only on the sample with spectroscopic
redshifts.  The solid line is the theoretical estimate of the Galaxy
{\it Baryonic} Mass Function, that is to be regarded as an upper limit
in$\Lambda$--CDM scenarios.  The dashed line is the GSMF predicted by
PLE models.  }
\label{massfun_obsteo1}
\end{figure}
the mass assembly of galaxies is driven by the merging of DM
halos. After initial collapse the latter coalesce to form larger and
larger structures (from galaxies and small groups to rich clusters of
galaxies), at a rate well approximated by the Extended Press \&
Schechter theory (Lacey \& Cole 1993, and references therein). 

Thus, galaxies are included into progressively larger and larger host
DM structures, where they may merge with the central dominant galaxy
due to dynamical friction, or undergo binary aggregations with other
galaxies orbiting the same host halo.  The analytic description of
these dynamical processes is bound by an a posteriori consistency with
high resolution N-body simulations.  Once these processes have been
fixed, the distributions of galaxies as a function of their DM
circular velocity can be computed without further ambiguity, and is
only slightly dependent on the assumptions used to describe the
dynamical processes involved. Assuming that baryons smoothly follow
the DM condensations, one can then derive the expected distributions
of the {\it available baryonic galaxy masses} that corresponds to the
mass distribution of the DM halos.  This is obviously an upper limit
to the actual GSMF. We have obtained such baryonic mass distributions
from the theoretical distribution of galaxy Dark Matter circular
velocities (from Menci et al. 2002, Fig.~3), by first computing the
corresponding DM mass as $M_{\rm DM} = v_c^3/10\,G\,H(z)$, where $v_c$
is the circular velocity of the galaxy DM halo and $H(z)$ is the
Hubble constant at redshift $z$, and then applying a simple scaling
$M_{\rm baryon} = M_{\rm DM} \frac{\Omega_{\rm baryon}}{\Omega_{\rm
DM}}$, with $\frac{\Omega_{\rm baryon}}{\Omega_{\rm
DM}}=\frac{0.045}{0.23}$ from the best-fit WMAP parameters (Bennett et
al. 2003), under the assumption that $\frac{\Omega_{\rm
baryon}}{\Omega_{\rm DM}}$ is constant down to galactic scales. As
discussed above, there are essentially no free parameters in deriving
these distributions, that depend primarily on the cosmological
parameters and on the dynamics of galactic sub-halos. These constitute
at present the most solid description of the hierarchical galaxy formation
models, so that the predicted distribution of $M_{\rm baryon}$ can be
considered as a solid upper bound to the observed GSMF.

The resulting distributions are then compared in
Fig.~\ref{massfun_obsteo1} with the empirical K20 stellar mass
distributions. The observed number density of massive
galaxies never exceeds this fundamental $\Lambda$--CDM constraint. This
is equivalent to say that at all explored redshifts there are enough
massive DM halos to account for the observed comoving number density
of massive (in $M_*$) galaxies (see also Gao et al.  2003).  There is
a tendency for the most massive galaxies at $z\lsim 1$ to be somewhat
close to this limit but, as discussed in Sect. 5,
%L% we will show later in the comparison with the PLE models, 
this may be due to
the impact of large scale structures in our sample.  
From Fig.~\ref{massfun_obsteo1} one derives  that massive galaxies have
already converted into stars $\sim 30-50\%$ of the available baryon
reservoir, and therefore have a ratio $M_{\rm DM}/M_*\simeq 10-20$,
consistent with the observed properties of local massive ellipticals
(Padmanabahn et al. 2003). This efficiency of baryon-to-star conversion
drops rapidly with decreasing mass of the host DM halo, in agreement
with the naive expectation that star formation can proceed to a higher
level of completion in deep potential wells, while early winds or other
effects easily evacuate of most of their baryons the less massive DM halos
with shallower potential wells. Several interesting ramifications may
follow from this semi-empirical estimate of the baryons-to-star conversion
efficiency as a function of the mass of DM halos, but following them in
any detail goes beyond the scope of the present paper. We just notice here
the relevance of this aspect for the chemical evolution of galaxies, the
IGM and the ICM, the run of the total mass to light ratio as a function
of galaxy mass, etc.

Another kind of constraint can be obtained from the Pure Luminosity
Evolution (PLE) models: by neglecting any merging, and requiring
consistency with the properties of present-day galaxies, these models
provide an upper limit on the stellar mass distribution within a given
cosmology.  We have estimated the GSMF evolution for the PLE case
starting with the local $K$-band luminosity function for the various
morphological types (Kochaneck et al. 2001), and by evolving back in
time the mass of each galaxy according to the e-folding time of the
star formation rate appropriate for the corresponding morphological
types (Pozzetti et al. 1996, 1998). The full procedure is described in
Paper V of this series (Pozzetti et al. 2003), and the resulting PLE
predictions are shown in Fig.~\ref{massfun_obsteo1}. At $z\leq 1$,
these predictions are in excellent agreement with the observed data,
with the possible exceptions of the more massive bins, where there
appears to be an excess of galaxies in the data with respect to the
PLE prediction. However, in the two bins around $z=0.7$ most of the
contribution to the top end of the GSMF comes from the prominent
concentrations (clusters) at $z=0.67$ and 0.73. 
At higher redshifts, the PLE
predictions are still well consistent with the data. At $z=1.5-2$, PLE
models are formally above the observed density by about 30\% around
$M_*\simeq 10^{11}M_\odot$ (see also Fig~\ref{mdens_teo}), where the
statistics is good and the results of the spectroscopic sample matches
the photometric one.

Overall, we conclude from this comparison that the evolution of the
massive galaxies in the K20 sample is consistent with the fundamental
constraints of the $\Lambda$--CDM scenario, and that the more massive
galaxies have already converted into stars a significant fraction of
their baryonic reservoir. The actual success (or lack of) encountered
by theoretical models in populating the various DM halos with the
observed amount of stars is discussed next.

%\newpage
%pippo
%\newpage
\subsection{Comparison with theoretical models of galaxy formation}

%\begin{figure}[ht]
%\resizebox{\hsize}{!}{\includegraphics{MF_teo_salp.ps}}
\begin{figure*}
\centering
\resizebox{\hsize}{!}{\includegraphics[angle=-90]{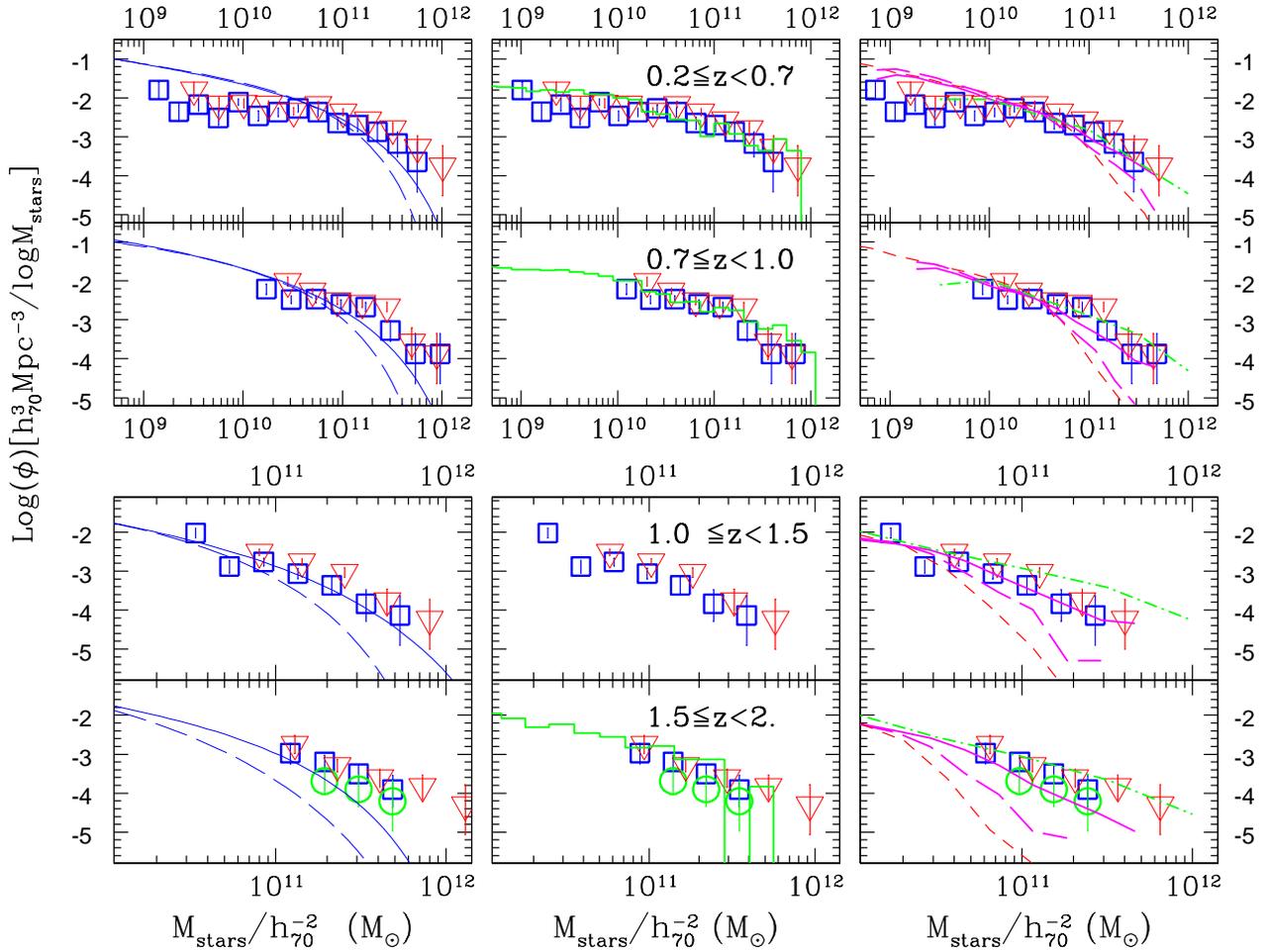}}
\caption{ Observed Galaxy Stellar Mass Functions in the K20 sample,
compared with theoretical rendition in a $\Lambda$--CDM hierarchical
Universe, divided according to the IMF adopted in the models. {\it
Left panel:} Salpeter (1955), {\it Central panel:} Gould et
al. (1996), {\it Right panel:} Kennicutt (1983). In the central and
right panels the observed GSMFs have been scaled to the corresponding
IMF as described in the text. The theoretical models are: {\it left
panel}: Menci et al. (2002) (dashed line) and Menci et al. (2004)
(solid line); {\it central panel:} Nagamine et al. (2001); {\it right
panel}: Cole et al. (2000) (short-dashed); Somerville et al. (2004a)
(S04 and S04b, thick solid and dashed lines); Granato et al. (2004)
(dot--dashed). See text for full details.  All the model predictions
are evaluated at the central redshift of each bin, with the exception
of the Nagamine et al. (2001) case which is computed at $z=0.5$, 1 and
2.}
\label{massfun_obsteo}
\end{figure*}

%It goes without saying that modeling 
%behavior of the baryonic component is far more uncertain than that of
%the DM component, hence it should be no surprise if necessarily
%simplified models such as the SAMs fail to reproduce the observations,
%even by a large margin.

Within the $\Lambda$--CDM scenario, several attempts have already been
made to model the history of star formation within DM halos, that
is far more uncertain than that of the DM component because of 
the highly non-linear behavior of the baryonic component.

A first class of theoretical models have been developed
using simple parametrized
prescriptions to relate the star-formation rate to the properties of
such halos, without attempting a physical modeling of the baryonic
component. These {\it semi-analytical models} (SAM) include e.g.,
those of Cole et al, (2000), Somerville et al. (2001), and Menci et
al. (2002,2004).

Another class of models are those complementing N-body simulations for
the dissipation-less components (DM and stars) with a full
hydro-dynamical description of the baryonic gas component: we shall
consider in the following the simulation of Nagamine et al. (2001a,b)
based on the Cen \& Ostriker (2000) models.

Finally, Granato et al.\ (2004) presented a semi-analytical type
model, focused on the relationship between the early formation of AGNs
and spheroids.

We shall proceed in this section to a comparison of the K20 results
with these theoretical models.  Fig.~\ref{massfun_obsteo} displays in
separate panels the direct comparison of the various model GSMFs with
their K20 counterparts.  As an alternative approach, we provide
another comparison with the theoretical models in
Fig.~\ref{mdens_teo}, where we plot the evolution the stellar mass
density contributed by galaxies in the range $5\times 10^{10} M_\odot
\leq M_* \leq 5\times 10^{11} M_\odot$ (i.e., around the Schechter
mass in the local mass function of Cole et al. 2001) which in the K20
sample can be computed with at most little extrapolations of the
GSMF. 

In making this exercise, we had to compensate for the different IMF
adopted in some of the theoretical models. This has been accomplished
by applying the appropriate scaling factor to our GSMF and stellar
mass densities, as described below.

\subsubsection{Semi-analytical models}

The main differences among the SAMs considered in this section
concern  the physical descriptions of processes such as 
the interactions among satellite galaxies orbiting the same host DM
halos (groups or clusters), the star formation processes during
galaxy interactions and merging events, the baryonic fraction, and 
the adopted stellar IMF. On the other hand, they all adopt the same
standard parametric laws for the ``quiescent'' star formation and the
supernova feedback.  In particular, we shall consider the following
SAMs:
\newline {\bf M02)} The SAM by Menci et al. (2002) with assumptions
largely similar to the C00 model.  The main difference is that
interactions among satellites are now considered to affect only the
mass distribution of galaxies when the orbital parameters are
conducive to bound mergers. It adopts $\Omega_b=0.03$, and a Salpeter
IMF.
\newline {\bf M04)} The ``interaction starburst'' model by Menci et al.
(2004): in addition to the processes considered in the M02 model, 
it includes the effects of fly-by galaxy interactions (not
leading to bound merging) as triggers for starbursts. The cross
section and the burst efficiency are taken from the physical model for
the destabilization of gas during galaxy encounters developed by
Cavaliere \& Vittorini (2000). This approach leads  to a higher average
contribution of  starbursts with respect to the S04a and S04b models
(see below).
\newline {\bf C00)} The SAM  of the Durham group, in the Cole et al. (2000)
rendition (see also Baugh et al. 2003).  In this model, the
interactions between satellite galaxies are not considered to affect
either the mass distribution of galaxies or their star formation in a direct
fashion. The model also adopts a low value of
$\Omega_b=0.02$ and a Kennicutt IMF.  
\newline {\bf S04a)} The ``merging starburst'' model of Somerville et al.
(1999, 2001) in its recent rendition (Somerville et al. 2004a), which includes 
the effects of merging between galaxies
on both their mass function and their star formation rate, the latter
being bursted in each merging event. The cross section and the burst
intensity are derived by extrapolating the results of hydro-dynamical
N-body simulations to the whole range of masses considered in the
model. In this model $\Omega_b$ is 0.04 and a
Kennicutt IMF has been adopted.
\newline {\bf S04b)} A ``reduced merging'' version of S04a, that has been
used to compare the predicted redshift distribution for $K_s\leq 20$ with
the K20 and GOODS data (Somerville et al. 2004b). Among the major
revisions, this ``reduced merging'' recipe has been adopted by
increasing the Dynamical Friction time and adding an additional time
delay for halo relaxation (see e.g. Mathis et al. 2002). In practice,
this model has the same star--formation recipes of S04a but a reduced
formation of massive objects.

To put both the latter three models and the K20 GSMFs on the same
foot, the observed masses have been systematically reduced by 0.35
dex. No such reduction is necessary when comparing with the M02 and
M04 models, since they were constructed adopting the Salpeter IMF.

In the low-redshift bin all the SAM mass functions appear to be
systematically steeper than the observed GSMF, with a pronounced
excess of low-mass galaxies, and an incipient deficit of massive
ones. The former discrepancy is a well known problem afflicting some
SAMs also at zero redshift (e.g., Cole et al. 2000; Baugh et
al. 2002), which is also directly noticeable in the comparison of the
luminosity functions from UV to IR (Poli et al 2001, Pozzetti et
al. 2003, Poli et al. 2003).

Fig.~\ref{massfun_obsteo} shows that most SAMs do reproduce the bulk
of the GSMF (i.e. the region around $M_*$), although they often under-produce
the very massive galaxies by an amount that increases with
redshifts. Moreover, the various models show a remarkably large spread
in the predicted GSMF, and progressively diverge from each other with
increasing redshift, reaching differences by more than 2 orders of
magnitudes at the highest explored redshifts.

\begin{figure}[ht]
\resizebox{\hsize}{!}{\includegraphics{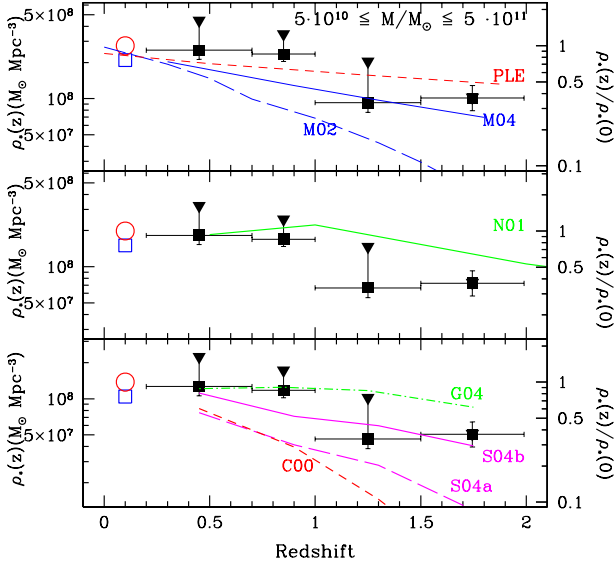}}
\caption{ Cosmological Stellar Mass Density for objects of $M_*\simeq
10^{11}M_\odot$ in the K20 survey, compared with the theoretical
models discussed in the text, divided according to the relevant IMF:
{\it upper panel}: Salpeter (1955), {\it central panel:} Gould et
al. (1996), {\it lower panel:} Kennicutt (1983).  Filled squares
represent the BF estimates, filled triangles the MA estimates. The
empty circle at $z=0.1$ represents the local value of Cole et al 2001,
the empty square the same value ``scaled'' to BF (see text for
details).  The models plotted are: Menci et al. 2002 (M02, long dashed
line), Menci et al. 2004 (M04, solid line) and PLE (short--dashed
line) in the upper panel; Nagamine et al. 2001 (N01, solid line) in
the central panel, and Cole et al. 2000 (C00, short-dashed);
Somerville et al. 2004a (S04a and S04b, thick solid and dashed lines);
Granato et al 2004 (dot--dashed line). See text for more details.
The relative value with respect to the local value of Cole et al. 2001
is shown in the right y--axis.  }
\label{mdens_teo}
\end{figure}

The same results can be obtained by looking at the evolution of the
stellar mass density of galaxies in the range $5\times 10^{10} M_\odot
\leq M_* \leq 5\times 10^{11} M_\odot$ (Fig.~\ref{mdens_teo}).  It is
worth noting that a significant decrease ($\sim 50\%$ from $z=0$ to
$z=2$) is exhibited also by the PLE models, despite the large $z_{\rm
form}=5.7$ adopted.

Given the complexity of the baryonic physics involved in galaxy
formation, all the parameterizations adopted by the various SAMs
appear a priori equally plausible. Yet, as shown by
Fig.~\ref{massfun_obsteo} and Fig.~\ref{mdens_teo}, the results
diverge dramatically with increasing redshift,
offering a rather powerful opportunity for the direct
observation of high redshift galaxies to discriminate between more or
less viable SAMs, thus possibly giving useful hints for a better
understanding of the dominant physical processes. For example, it
appears that merging- or interaction-induced starbursts may be an
essential ingredient in order to produce a number of massive galaxies
at $z>1$, approaching the corresponding number that this survey has
detected.  On the other hand, the standard, ''quiescent'' star
formation regulated by the cooling time of baryons within the DM halos
(as adopted e.g., in C00 and M02) appears to produce the most
discrepant results compared to the observations presented in this
paper.  Also, the comparison suggests that the standard recipes for
dynamical friction adopted by S04b and M04 tend to perform better than
the ``reduced'' merging treatment (S04a), that was  introduced to
improve the  match with the density of bright galaxies in the local
universe.

%\newpage
%~
%\newpage

\
\subsubsection{N-body/hydrodynamical models of galaxy formation}

Unlike SAMs, the Nagamine et al. (2001a,b) simulation dynamically
solves a full set of fluid equations for baryons, and includes
physical processes such as radiative cooling and heating of the gas,
star formation, and supernova feedback, with star particles being
created out of the gas where it is contracting and cooling rapidly. To
some extent, the effect on star formation of galaxy interactions is
also automatically included in these simulations.

The published versions of these models provide GSMFs only for redshift
bins centered at $z=0.5$, 1, and 2, that we compare in
Fig.~\ref{massfun_obsteo} with the empirical K20 GSMFs in the
$0.2<z<0.7$, $0.7<z<1.0$ and $1.5<z<2$ redshift bins,
respectively. Since these models adopt the IMF from Gould et
al. (1996), the K20 masses have been systematically scaled down by a
factor 0.72, which is the appropriate correction to take into account
to convert them to a Salpeter IMF (Nagamine, private communication).

Fig.~\ref{massfun_obsteo} shows that these models predict an evolution
of the GSMF that is in very good agreement with the K20 results,
avoiding both the low-mass excess and the high-mass deficit typical of
SAMs. The good agreement in the low-mass range is likely to be due to
the stronger feedback effect that in these models is tuned to match
the X-ray background, resulting from the AGN activity. %In this
%context, it is worth noting that a strong contribution of AGNs to the
%feedback was also advocated by Granato et al. (2001, 2004) in order to
%suppress star formation in low-mass halos at early times.

Fig.~\ref{mdens_teo} shows that the Nagamine et al. models also
predict a very mild decrease with $z$ of the high-mass contribution to
the stellar mass density, actually even milder than that derived from
the K20 data.  We note also that these models predict that $\sim 30\%$
of the stellar mass is already in place by $z=3$ (Nagamine et
al. 2004), in agreement with the expectation from the so-called
``fossil evidence'' (Renzini 1998, 2003).

\subsubsection{Models for the joint evolution of QSO and spheroidal
 galaxies.}

Yet another class of models has been developed specifically to explore
the mutual feed-back between star formation in spheroids and the
high-z QSO activity (Granato et al.\ 2001, 2004), largely neglected by
classical SAM. In the latter paper, dark matter halos form at the rate
predicted by the canonical hierarchical clustering scenario, and
processes such as collapse, heating, cooling and supernovae feed-back
are taken into account with techniques and recipes typical of other
SAM. However, (i) it is assumed that angular momentum plays a
negligible role in slowing down star formation activity in massive
halos virialized at high redshift, which would be those leading to
spheroid formation and are the only one accounted for in this model;
(ii) the growth by accretion of a central SMBH is promoted by the
resulting huge, and heavily obscured, star formation rate and (iii) the
feed-back on the ISM due to the ensuing AGN activity is taken into
account with recipes inspired by the physics of Broad Absorption Line
QSOs.

This model adopted a double--slope IMF that provides a \ml\ ratio very
close to Kennicutt. For this reason, we will plot this curve in the
same panel with the Kennicutt models. We note that in the case of the
second redshift bin ($0.7<z<1$), the model has been computed at $z=1$,
and should therefore be considered as slightly underestimated.

The comparison with the K20 data suggests a good agreement in the
$z\leq 1$ range, and an overestimate of a factor about 2 of the mass
density of massive galaxies at $z>1$. In this redshift range, this
models predicts a fraction of massive galaxies to be strongly
extincted, and therefore to escape our $K<20$ selection criteria:
further modeling is required to assess whether this effect is able to
reconcile the model with our observations.

\section{Summary and Discussion}
In this work, we have used the spectroscopic redshifts and the
spectrophotometric properties of the galaxies in the K20 sample to
estimate their stellar masses and to build the corresponding GSMF at
different redshifts.  Our basic results, that are based on the
assumption of a Salpeter IMF, can be summarized as follows:

$\bullet$ We have used two different methods to estimate the stellar
mass $M_*$ and the rest-frame luminosities for each galaxy in the
sample.  In one case, we assume all galaxies started to form stars at
$z=20$ with a star-formation rate declining exponentially
thereafter. The e-folding time is then determined by demanding the
stellar population model to reproduce the observed $R-K$ color.  We
referred to this approach as the ``Maximal Age'' (MA) method. In the
other approach, we remove any constraint on age, metallicity and dust
content, but require consistency with the whole multicolor spectral
energy distribution.  We referred to this approach as the ``Best Fit''
(BF) method.

$\bullet$ Wehave performed a comparison of the two methods, showing
that the typical BF masses are lower with respect to the MA estimates
by a factor that is typically $\simeq 2$ for objects of
$M_*=10^{10}M_\odot$ and $\simeq 1.5$ for objects of
$M_*=10^{11}M_\odot$.  This lower estimate is due to the lower galaxy
age obtained on average from the BF procedure.  We have also carefully
inspected for systematic effects by means of simulations, comparison
with the spectroscopic information and by looking at the intrinsic
degeneracy among the input parameters.  We conclude that the two
methods together should give a fair representation of the existing
uncertainties in the derived masses.

$\bullet$ The final galaxy sample of the K20 survey spans a range of
stellar masses from $M_*=10^9M_\odot$, at the lowest redshifts, to
masses close to $M_*=10^{12}M_\odot$.  
Such massive galaxies appear to be common at $0.5<z<1$, and are detected 
up to $z\simeq 2$.

$\bullet$ With the K20 data set, we have built the Galaxy Stellar Mass
Function (GSMF) and the corresponding total mass density in four
redshift bins centered at $z=0.45,0.85,1.25,1.75$, while introducing a
correction to take into account for the incomplete coverage of faint
objects with high \ml.

$\bullet$ Up to $z\simeq 1$, we observe only {\it a mild evolution of
the GSMF and of the corresponding global stellar mass density}.  Both
the MA and the BF estimated GSMFs indicate a decrease by $\sim
20-30\%$ of the number density of objects around $10^{11}M_\odot$
(where the statistics is sufficiently accurate).  This implies that
the evolution of objects with mass close to the local characteristic
mass is essentially complete by $z\simeq 1$.

$\bullet$ At higher redshifts a drop begins to appear in the comoving
number density of galaxies within the explored mass range,
corresponding to a decrease in the normalization of the GSMF.  At
$z\simeq 2$, a fraction of $\sim 30-40\%$ of the present day stellar
mass in objects with $ 5\times 10^{10}M_\odot < M_* < 5 \times
10^{11}M_\odot$ appears to be in place.

$\bullet$ We detect a change in the physical nature of the most
massive galaxies: at $z \lsim 0.7$, all galaxies with
$M>10^{11}M_\odot$ are either early or early$+$emission type, while
the mass density due to massive star--forming galaxies increases with
$z$: in the highest redshift bin, we estimate a lower limit (due to
incomplete spectroscopic identification) of 21\% to their contribution
to the observed stellar mass density.

$\bullet$ The observed evolution of the \mlr \ ratio (Sect. 4.2)
provides evidence for a differential evolution of early type galaxies,
suggesting that more luminous (i.e. more massive) galaxies appear to
reach near completion first, while less massive ones keep growing in
mass till later times. A direct detection of this effect in our GSMF
is hampered by the low statistics, although we note that there is
a tentative indication that the decrease with redshift of the GSMF is
more pronounced towards the low-mass end of the explored range than
for the most massive objects.  Clearly, much wider surveys are
required to confirm this potentially crucial item.

$\bullet$ The global rise of the stellar mass density from $z\simeq 3$
to $z\simeq 1$ is broadly consistent with the integrated contribution
from the global star--formation as inferred from UV--selected
galaxies, once a modest amount of dust extinction (E(B-V)=0.15) is
accounted for.

$\bullet$ It is shown that the large number of massive galaxies
detected at high $z$ does not violate any fundamental $\Lambda$--CDM
constraint. Specifically, up to $z=2$ there is no shortage of DM halos
massive enough to account for the baryonic mass of the observed
galaxies. Very interestingly, the fraction of baryons converted into
stars appears to strongly increase with the mass of the host DM halo.

$\bullet$ We have compared in some detail these results with the
expectations of updated models for galaxy formation in a
$\Lambda$--CDM Universe, including several {\it ab initio} renditions,
either semi-analytical (namely: Cole et al. 2000; Somerville et
al. 2004a,b;; Menci et al. 2002, 2004) or hydrodynamical simulations
(Nagamine et al. 2001a,b), as well as the physically motivated model
of joint evolution of QSO and galaxies by Granato et al. (2001,2004).
The predicted evolution of the GSMF varies quite dramatically from one
rendition to another, being very sensitive to model ingredients such
as interaction-driven starbursts, feedback, etc.  Some semi-analytical
models are consistent with the observations up to $z\simeq 1.5$, and
slightly underestimates those at higher $z$, while other underpredict,
in some case by a large factor, the number density of massive galaxies
at high redshift. Conversely, the hydrodynamical simulations of
Nagamine et al. (2001a,b) and the Granato et al. (2004) models appear
to match the observed number density at $z\simeq 1$, where they equal
or even exceed the PLE predictions, and even overpredict the ones up
to $z\simeq 2$.  It is worth noting that the strong contribution of
AGNs to the feedback in the models of Nagamine et al. (2001a,b) and
Granato et al. (2001, 2004) is apparently effective in order to
suppress star formation in low-mass halos at early times, and to
reproduce the slope of the GSMF at low masses.  On a different line,
the PLE predictions appear to be consistent with the observed data, at
least up to $z=1.5$.

$\bullet$ We have accompanied all these findings by several
cautionary remarks concerning their sensitivity to the adopted IMF,
and possible biases due to surface brightness dimming effects, highly
obscured objects, the narrowing range of explored masses at high
redshift, and cosmic variance.

\smallskip

While keeping these caveats in mind, the present results allow to
sketch a global scenario for the evolution of massive galaxies.  Up to
$z\simeq 1$, clearly little evolution has taken place in these
objects. Their number density is close to the local one ($\sim 70-80\%$)
and to the prediction of simple PLE models. In addition, most of the
mass density resides in early-type, passively evolving galaxies, that
must have formed the bulk of their stars at least $2-3$ Gyr before,
i.e., at $z\simgt 2$. In this respect, the K20 survey strengthens
similar conclusions by Brinchmann \& Ellis (2000), using the CFRS data
(Lilly et al. 1995), by D03 and F03 in the much smaller HDF
fields. The stability of the massive, early type
galaxy population up to this redshift is also recovered by the COMBO17
Survey (Bell et al.  2004), accompanied by a progressive
disappearance of the less massive early-type (red) galaxies.

%This apparently contrasts with the results of the MUNICH
%survey (Drory et al. 2001), but the lack of spectroscopic redshifts
%and the different technique adopted to estimate the stellar mass makes
%such a comparison difficult. 

Beyond $z\sim 1$ the evolution of massive objects starts to accelerate
and by $z\simeq 1.8$ $\sim 30-40\%$ of the local density has been
locked in massive galaxies. This is associated with a change in the
physical properties of massive galaxies, among which star-forming
objects now become common, and contribute a much more substantial
fraction of the observed mass density (at least 21\% in the K20
sample). So, several massive $K<20$ galaxies in the high-redshift tail
($z\geq 1.7$) of the redshift distribution are star-bursting objects
showing irregular morphologies, and may be strongly clustered (Daddi
et al. 2004a), which makes them likely progenitors of local massive
early-type galaxies. While up to $z\sim 1.5$ PLE models still give an
acceptable fit to the data, at higher redshifts the massive starbursts
in the sample mark a departure from such models, and signal that one
may be entering the formation epoch of massive spheroids (Daddi et
al. 2004; Somerville et al. 2004b). However, with the K20 survey we
have got just a first glimpse to the transition from the
passively-evolving to the active star-forming progenitors of the local
early-type galaxies.  Wider areas and deeper spectroscopic surveys are
clearly required to thoroughly map this transition, that may extend up
to $z\sim 3$ and beyond.

Overall, these features are in agreement with the basic elements of
the $\Lambda$--CDM scenarios. In an accelerating universe, massive
galaxies have time to complete their assembly at $z\simeq 1$, avoiding
too much merging at low redshifts. Thanks to the availability of gas,
shorter cooling time, and higher interaction and merging rates star
formation naturally occurs most efficiently at earlier times.  The
anti-correlation of feedback with the halo mass, and the correlation
between high density peaks, boosts the star--formation rate in high
mass objects at high $z$, with respect to lower mass ones. Despite
these ``built in'' features, most renditions of the $\Lambda$--CDM
paradigm tend to delay the star formation and assembly of massive
galaxies well beyond the epochs favored by the present observations,
and several of them significantly fail to match the observed mass
densities.  The reason is tied to the different physical processes
that are contained - or ignored - in the models and to the way they
are described. 

At the time of writing, the Spitzer satellite is acquiring its
first data: it is easy to predict that it will bring new, decisive
information in this critical issue.

The \ml ratio, stellar mass densities and GSMFs described in this
papers are available in electronic form at the web site {\sf
http://www.arcetri.astro.it/$^{\sim}$k20}

\begin{acknowledgements}
We warmly thank Carlton Baugh, G.L. Granato, Ken Nagamine, and Rachel
Somerville for providing files with the predictions of their models,
A. Grazian for a careful reading of the manuscript and precious
suggestions, and the anonymous referee for very constructive comments
that resulted in a significant improvement of the paper.  We thank the
VLT support astronomers for their kind assistance during the
observations.

\end{acknowledgements}

\bigskip

\noindent
{\bf {\large Appendix A: Validating the ``Best Fit'' method}}

\bigskip

{\it {\large A.1 Comparison between the Maximal Mass and Best Fit mass estimates}}
\smallskip

\begin{figure}[ht]
\resizebox{\hsize}{!}{\includegraphics{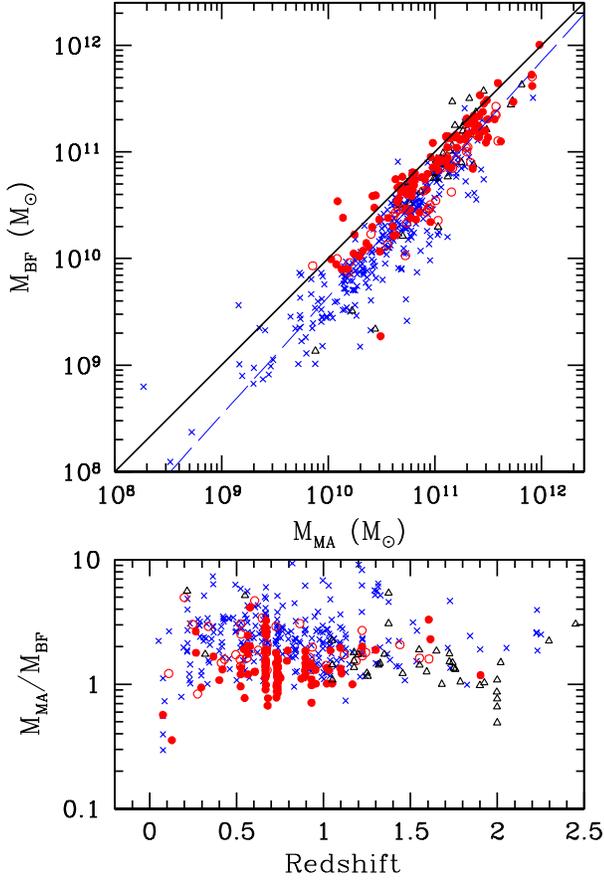}}
\caption{Comparison between the stellar masses estimated for the K20
galaxies with the Maximal Age and Best Fit techniques. {\it Upper
panel:} The relation between MA (x--axis) and BF (y--axis) stellar
masses in the K20 sample with full multicolor coverage.  Different
symbol refer to spectroscopic classification, as in Fig~\ref{mass_z}: 
solid circles, early type; empty circles, early+emission type; crosses, late type; 
triangles, photometric redshifts.
The dashed line shows the first--order fit to the observed
relation. {\it Lower panel:} Ratio between the MA and BF sample as a
function of the redshift, for the same sample. Symbols have the same
meaning of the upper panel.}
\label{MM_BF}
\end{figure}

Figure~\ref{MM_BF}  shows the comparison between the
stellar masses estimated with the two different criteria.  Beyond the
obvious correlation, it is immediately clear that the MA models
provide an estimate that is typically larger than the BF estimates,
although a large scatter exist. The ratio between the two methods is
larger at intermediate masses, and it is on average a factor
$<M_{\rm MA}/M_{\rm BF}> \simeq 2$ at $10 < log(M_{\rm BF}/M\odot) < 10.6$, and
converges toward a factor $<M_{\rm MA}/M_{\rm BF}> \simeq 1.6$ at
$log(M_{\rm BF}/M\odot) > 10.6$.  We have found that this difference is
mainly due to the lower average ages that are inferred for the K20
galaxies in the BF approach (the median $z_{\rm form}$ resulting from the ``BF'' 
approach is about 2).
%L% On the one side, 
We have verified that
this trend still holds if we fix the metallicity to a solar value, or
if we use different extinction curves, like e.g. Calzetti 2000. 
%L% On the other side, we have found that the mass difference is strongly
%L% correlated with the age difference between the two estimates (the
%L% median $z_{form}$ resulting from the ``BF'' approach is about 2). 
A few BF models have larger stellar masses than the corresponding MA
estimates: these results from objects fitted with large ages
(comparable with the MA assumptions) and with combinations of dust
and/or metallicities that further enhance the \ml ratio. We also note
that a contribution to the scatter (with a 10\% r.m.s.) results from
the different normalizations (see Sect. 3.3).

We have obtained a simple first order regression to the points in
Fig.~\ref{MM_BF}, finding that the relation $log M_{\rm BF} = 1.106 (\pm
0.001) \times (log M_{\rm MA}) -1.42 (\pm 0.06)$ provides a reasonable fit
to the observed relation, accurate to $\Delta log M_* \simeq 0.14$. 
%L% We have used this relation to obtain the BF mass estimates of the 72
%L% objects in the CDFS field (all at $z\leq 1.5$) for which we do not
%L% have a full multicolor coverage.

%In summary, we have found that there is a main physical difference
%between the results drawn from MA and BF models, since BF models
%typically predict galaxy ages much shorter than MA models, where
%$z_{form}=20$ has been assumed for all galaxy types.  As a
%result, the stellar masses estimated from the BF models are
%systematically lower than those of the MA models, by an amount that is
%typically a factor of 2 at $M_* \simeq 10^{10} M_\odot$ and a factor
%50\% at $M_* > 8 \times 10^{10} M_\odot$.

%In the lower panel of Figure~\ref{MM_BF} we plot the ratio between the
%MA and BF mass estimates as a function of redshift. Although it is
%suggested a mild anti-correlation of $M_{MA}/M_{BF}$ with $z$, our
%statistics at large $z$ is too scanty, and limited to the highest
%masses, to establish this quantitatively. We note in particular that
%the two estimates appear to be consistent at $1.5<z<2$, where our
%highest redshift bin will be computed, making this estimate
%particularly well constrained. With more scatter (and poor
%statistics), objects at very low $z$ also appear to be converge,
%consistent with our analysis of the local SDSS sample (Appendix A.5).

\bigskip

\noindent
{\it {\large A.2 The effect of secondary bursts}}

\smallskip

\begin{figure}[ht]
\resizebox{\hsize}{!}{\includegraphics{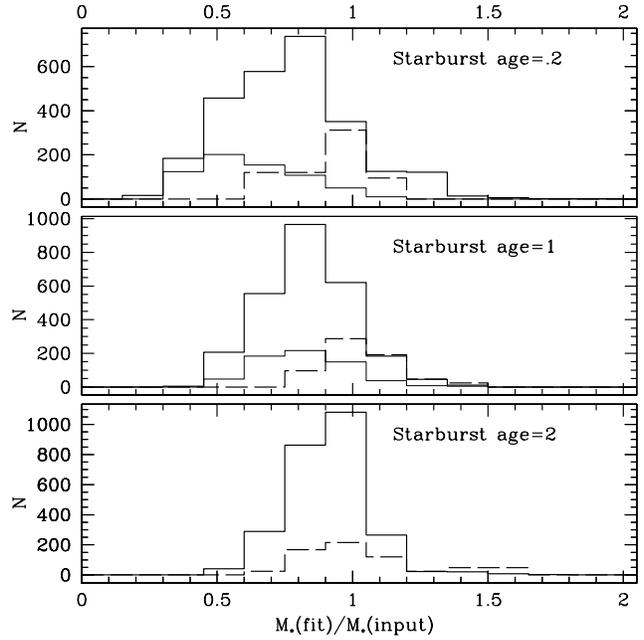}}
\caption{Results of the simulations performed to check the reliability
of the Best Fit stellar mass estimates  in the cases of star--formation
histories characterized by multiple starbursts. The histogram shows the 
ratio between the fitted mass $M_{\rm fit}$ and the input one
$M_{\rm input}$ at three different ages of the second starburst. The solid line
shows the  results of the
simulations done assuming that the second starburst assembles
a fraction $f$ from 10\% to 70\% of the total mass. Thin solid line is the 
simulation with $f=10$\%. Dashed line shows a simulation with the starburst component 
only.
}
\label{simul}
\end{figure}

In order to check the reliability of the BF stellar masses estimates,
in particular during secondary starbursts, we have performed a set of 
simulations.
We have computed a mock multicolor sample assuming a scenario where
galaxies were built during two major starbursts. During the first one,
characterized by an initial redshift $z_{\rm form1}$ and exponential
timescale $\tau_1$, a fraction $1-f$ of the total final mass is
assembled. All possible permutations of $z_{\rm form1}=10,8,6,4,3,2.5$,
$\tau_1=0.1,0.3,1,5$~Gyrs were adopted, to ensure a proper sampling of
any spectral type, including very red, passively evolving
galaxies. Solar metallicity and no dust was assumed in this case.  In
addition, a younger starburst was summed, contributing for a fraction
$f$ of the final mass, with star--formation time scales
$\tau_2=0.1,0.3,1$~Gyrs and ages $0.1,0.2,..2$~Gyrs.  Solar
metallicity and $E(B-V)=0.2$ was adopted. The resulting mock spectral
distributions were computed with $f=0.1,0.3,0.5,0.7$ at
$z=0.5,0.6...1.3$ to cover the main range of our observations, and
were analyzed with the same recipe of the K20 sample. In
Fig\ref{simul} we show the histogram of the recovered stellar mass
$M_{\rm fit}$ with respect to the input one $M_{\rm input}$, at different
starburst ages. It is shown that if the second burst is caught during
its early phase, within the first $0.2-0.3$ Gyrs from its start, the
BF estimated mass is typically lower than the actual one by a factor
of about 25\% (average value). We have found that the effect is
larger when the fraction $f$ is small (as shown from the
$f=0.1$ case in the histogram of Fig\ref{simul}) and/or the first
starburst is peaked at high $z$.  At larger ages, the BF estimated
mass approaches the input one, leading to a final average
underestimates of 20\% at a starburst age of 1 Gyrs and only 10\%
after 2Gyr. As a check,we have also verified that if we include only
single exponential laws (dashed line in Fig.~\ref{simul}) the resulting
fit is essentially unbiased.

These simulation show that if the observed galaxy have a
star--formation history significantly departing from single
exponential, and are observed during a major starburst, the overall
effect on the estimate of the stellar mass is relatively small,
especially for the reddest galaxies, that dominate the massive tail of
the GSMF.

\bigskip

\noindent
{\it {\large A.3 The D4000 break}}

\smallskip

\begin{figure}[ht]
\resizebox{\hsize}{!}{\includegraphics{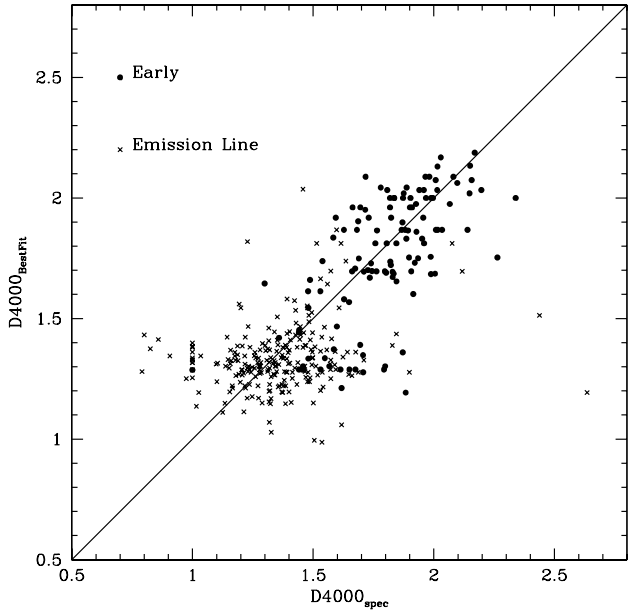}}
\caption{ Comparison between the D4000 break measured from the K20
spectroscopic observations (x axis) and the D4000 break estimated from
the Best Fit models (y axis). Filled dots represent late--type
galaxies while crosses the star--forming ones.}
\label{D4000}
\end{figure}
The $D4000$ break is known to be a sensitive probe of the age of the
underlying stellar population, and as such is a precious information
to test the star formation scenario that is produced by the BF models.

We plot in Fig.~\ref{D4000} the comparison between the $D4000$ break
measured from the K20 spectroscopic observations and the $D4000$ break
estimated from the Best Fit models. The comparison shows a relative
good agreement, with a nearly null average difference (that is $\Delta
D4000=0.02$) and a dispersion of 0.2, that is smaller than the
average error on the spectroscopic $D4000$. Fig.~\ref{D4000} shows
that, as expected, the $D4000$ break of early--type galaxies is pretty
larger than for later types.  In particular, we have found that the
average $D4000$ of the ``old'' EROs population, derived from BF models,
is very close to the
value of the average spectrum, that is 1.9 (Cimatti et al. 2002c).

\bigskip

\noindent
{\it {\large A.4 Quantifying the internal degeneracies}}

\smallskip

\begin{figure}[ht]
\resizebox{\hsize}{!}{\includegraphics{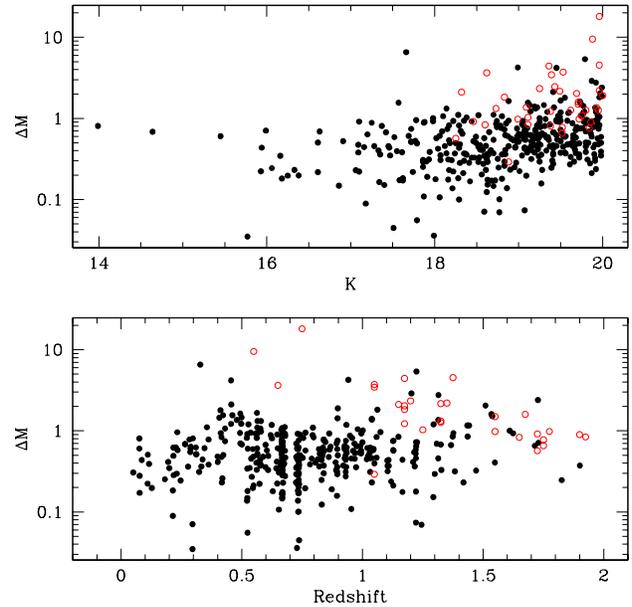}}
\caption{Confidence levels on the estimated BF mass, as a function
of $K$--band magnitude (upper panel) and redshift (lower panel).
The confidence level $\Delta M_*$ 
is defined as $\Delta M_* = (M_{\rm max} - M_{\rm min})/2 M_{\rm best}$
Filled points represent objects with spectroscopic redshift, empty with
photometric redshifts.
}
\label{error}
\end{figure}

Given the wide range of the free parameters involved in the BF
estimate, and the relatively loose constraints that can be obtained
from broad band imaging only, the best fit solution to each galaxy is
far from being univocal.  An advantage of the $\chi^2$ approach is
that it allows to take into account the resulting degeneracies among
the input parameters adopted and to provide an estimate of the range
of ``acceptable'' models.  At this purpose, we use a technique already
adopted in F03, and similar to the one adopted by
Papovich et al. 2001, based on the reduced chi--square $\chi^2$ (as
computed in Fontana et al. 2000).  The $1\sigma$ confidence levels on
the fitted parameters (such as mass, age and star--formation rate)
have been obtained by scanning the model grid and retaining only the
models that have $\chi^2\leq\chi^2_{\rm bestfit}+1$.  Prior to this, as in
Papovich et al. 2001, we have rescaled the noise in bright objects order to have
$\chi^2_{\rm bestfit}=1$ The scan is performed either at fixed redshift
(for objects with known spectroscopic redshift) or allowing the models
to move around the best--fitting photometric redshift.

%We scan all the spectral synthesis models of Tab\ref{tabBC}
%retaining only those that have a probability $P(\chi^2,n)\geq p_{th}$,
%where $p_{th}$ is a threshold that defines all the ``acceptable''
%models. To provide a robust estimate of the overall uncertainties, we
%have scaled all the errors in magnitude in order to have $\chi^2 = 1$
%for the best fit solution, and adopted a threshold $p_{th} = 0.05$.
In general, the range of acceptable values will not be symmetric with
respect to the best--fit solution: we find that the typical range for
a galaxy of stellar mass $M_*$ is $0.6 M_* - 1.6M_*$.  This
uncertainty is in agreement with the similar results in the HDFN
(D03) and HDFS (\cite{fontana2003}) at the same
redshifts and signals a core level of degeneracy in the input models,
that cannot be resolved without detailed spectroscopic informations.
This level is lower than the comparable estimates at $z\simeq 3$
(\cite{Papovich2001}, \cite{shapley}), since we can rely on a better
sampling of the NIR side of the spectrum.

To show how the uncertainty depends on the redshift and observed flux,
we use a simplified (symmetrical) estimator computed as $\Delta M_* =
(M_{\rm max} - M_{\rm min})/2M_{\rm best}$, that we plot as a function of redshift
and $K$--band magnitude in Fig.~\ref{error}.  For most of the objects,
this estimator is below 1, consistent with the average uncertainty
quoted above, showing that our mass estimates are overall robust
within a factor of 2.  As expected, faintest objects have a typically
larger uncertainty, as well as objects with $z>1.5$, that begin to
suffer from the poor sampling of the IR side of the spectrum: longer
wavelengths data, as those provided by the recently launched
Spitzer satellite, are required to improve their estimates.

Finally, we have found that MA model typically lay close to the upper
confidence level. Quantitatively, we have found that about
40\% of the MA values are lower than the upper $2\sigma$ confidence
level, and that the MA values are on average only 10\% larger than the
upper $2\sigma$ confidence level.

\bigskip

\noindent
{\it {\large A.5 The local sample}}

\smallskip

We address here the impact that our two different techniques to
estimate the stellar masses have on the estimate of the local stellar
mass functions and density ( \cite{cole2001}), which are the pivots of
the evolutionary trends that we have analyzed.  At this purpose, we
have followed the procedure described by \cite{bell} to build a sample
of local galaxies (i.e. $z\simeq 0.1$) by cross-correlating the SDSS
and 2MASS public catalogs. Using a sub area of the SDSS EDR, we
obtained a catalog of 6332 galaxies with full $ugrzJHK$ photometry,
from which we obtained the MA and BF stellar masses as described in
the main text.  The main results of this comparisons are:

\noindent
- MA stellar mass estimates do not depend sensitively on the color
adopted.  We have tested that using either the $J-K$, as done by
\cite{cole2001}, or the $R-K$ (as we did in our sample) or the $G-J$
(that grossly mimics the $R-K$ of our sample) the MA estimates do not
change systematically.

\begin{figure}[ht]
\resizebox{\hsize}{!}{\includegraphics{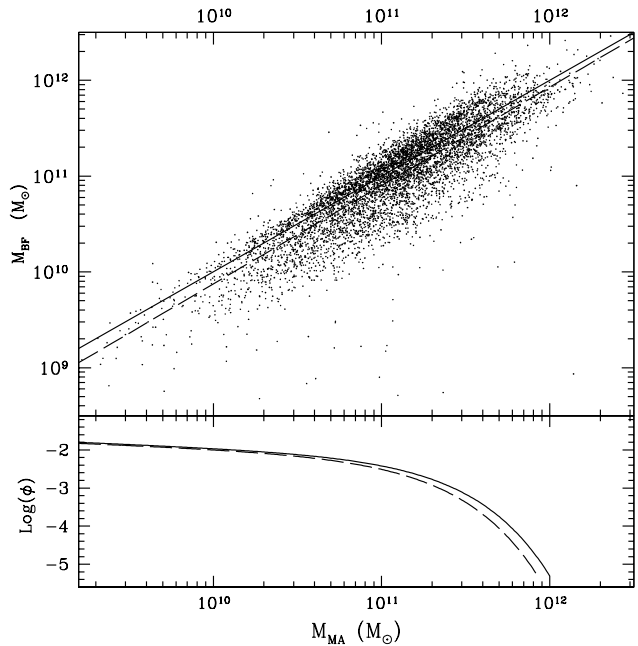}}
\caption{A comparison between the stellar masses estimated with the
Maximal Age and with the Best Fit methods on 6500 galaxies at $z\simeq
0.1$ taken from the SDSS and 2MASS surveys. Solid line is the
$M_{\rm BF}=M_{\rm MA}$ locus, dashed line is the linear regression among the
observed points.
}
\label{localM}
\end{figure}

\noindent
- In the local Universe, BF stellar estimates are systematically lower
than MA by about 20\%, with a trend of decreasing offset for more
massive objects. This is shown in fig\ref{localM}, where the two
samples are compared. A linear regression between $M_{\rm MA}$ and
$M_{\rm BF}$ yields $M_{\rm BF}=1.027 \times M_{\rm MA} -0.3955$.

Using the above relation to statistically convert the \cite{cole2001}
mass function to a BF one, we obtain the stellar mass function shown
in the lower panel of fig\ref{localM}, that marginally departs from
the original of \cite{cole2001} in the massive tail. Whenever we have
analyzed the evolution of the mass density with redshift with our
``BF'' estimates, we have adopted this ``BF--scaled'' local mass
function to compute the corresponding local mass densities.

\bigskip

\noindent
{\bf {\large Appendix B: Incompleteness effects in the stellar mass function}}

\smallskip

\begin{figure}[ht]
\resizebox{\hsize}{!}{\includegraphics{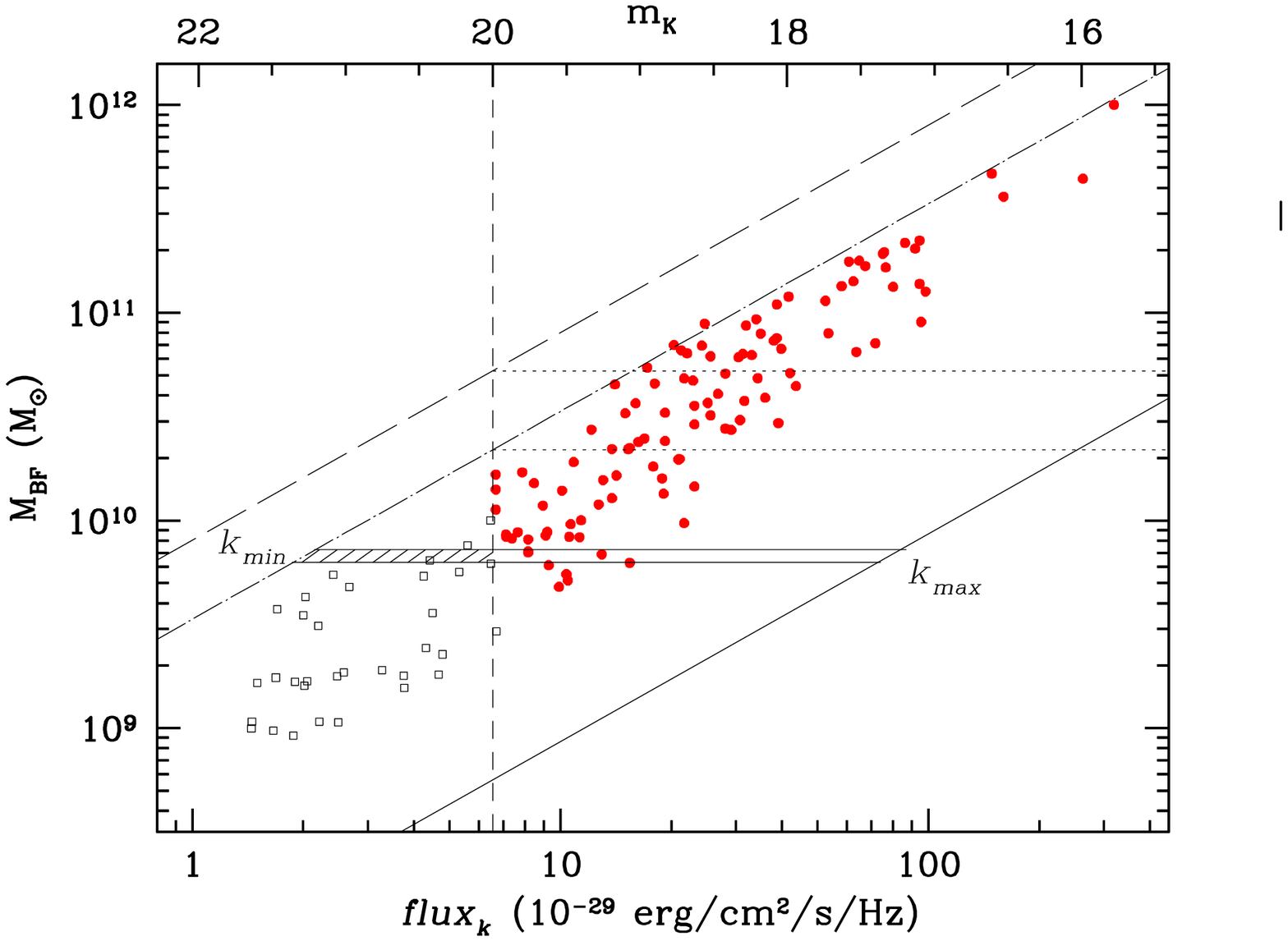}}
\caption{ Relation between the observed $K_s$ flux (lower scale; in
the upper scale, the corresponding $K_s$ magnitude) and the BF
estimated stellar masses, at $z\simeq 0.7$. Filled points are taken
from the K20 sample, empty are taken from the HDFS sample
(\cite{fontana2003}) for comparison only.  The diagonal lines bracket
the whole range of masses that are allowed to galaxies of the
corresponding $K_s$ magnitude in the BF model grid.  The solid line
show the minimal mass for objects of given flux. The dashed line is
the maximal mass, if one takes into account dusty objects, while the
dashed--dotted line is the maximal mass for dust-free, passively
evolving objects.  The dashed vertical line shows the $K_s<20$ limit
of the K20 survey.  The dotted line show the resulting (strict)
completeness limit on stellar masses, according to the
selection curve adopted. The shaded area shows the fraction of galaxies
lost (at a given mass) by incomplete coverage of the \ml ratio.  }
\label{complet}
\end{figure}

As discussed in the main text, the building of the corresponding
stellar mass function follows the traditional techniques ($1/V_{\rm max}$
and Maximal Likelihood) used for luminosity functions, although a
treatment must be included to correct for the incompleteness (in mass)
at the faintest levels, that arise from the not univocal conversion
from the observed K-band to the stellar mass.  To emphasize this
issue, we plot in figure~\ref{complet} the stellar masses
$M_{\rm BF}$ obtained in the K20 sample at the redshift of $z\simeq 0.7$
as a function of the observed $K_s$ flux. The K20 sample is shown
with filled points. To ease the visual representation of the
incompleteness effect, we add with empty symbols the corresponding
points at $K_s>20$ from a similar analysis carried on in the HDFS
(F03).

The relation follows the expected correlation with a scatter of about
$\Delta log(M_*)\simeq 0.6$, due to the intrinsic scatter in the
$M_*/L_K$ ratios. We note that the $K_s=20$ limit cuts the galaxy
strip at a mass level that varies from $\log (M_*/M_\odot) = 9.5$ to
$\log (M_*/M_\odot) = 10.3$.  The incompleteness at the faint $M_*$
levels becomes evident by observing the point distribution at a fixed
$M_*$ level.  At, say, $M_* \simeq 8 \times 10^{9}M_\odot$, about half
of the galaxies lay at $K_s>20$, and are therefore missed by a
$K_s\leq 20$ sample.  To recover from this incompleteness, one can
make use of the distribution of the \ml ratio, both as expected from
the library adopted and as observed in the data, as we describe below.

By construction, the points must lay between the minimum and maximum
mass that a galaxy at $z\simeq 0.7$ may have at a given $K_s$
magnitude: these limits can be obtained from the model grid that is
assumed to describe the galaxy properties.  In the case of the BF
grid, we have computed these thresholds by scanning the output of the
spectral library defined in Table~\ref{tabBC}, which, as discussed in
the text, is particularly sensitive to the effects of dust. We plot in
Fig.~\ref{complet} both the upper limit corresponding to dusty objects
(dashed line) as well as the limit corresponding to passively evolving
dust--free models (dashed-dotted line). The lower limit (solid line)
is in both case resulting from a star--forming, dust-free young
population.

Strictly speaking, the sample is therefore complete (in stellar mass)
only down to the horizontal dotted lines shown in Fig.~\ref{complet},
that {\it correspond to the maximal mass allowable (at any redshift)
for a galaxy of magnitude equal to the faint limit of the sample}.  In
principle, stellar mass functions and other quantities should be
computed only on the subsample with $M_*(z)\geq M^*_{\rm compl}(z)$. At
masses smaller than this limit, any magnitude--selected sample will be
progressively incomplete, although objects of \ml will still be
detected.  We note that this effect has not been taken into account in
the local estimates of the GSMF (\cite{cole2001}, \cite{bell}), which
may explain the decrease of the local GSMF that is observed in their
faintest bins.

However, as shown in Fig.~\ref{complet}, the actual threshold
sensitively depends on the choice of this ``maximal mass'' model, and
the strict adoption of a selection criteria would result in the loss
of a significant fraction of our sample: for these reasons, we have
introduced a correction for incompleteness that both allows to recover
a significant fraction of the sample and, using the observed
distribution of \ml, removes the critical dependence on the upper and
lower limits.

We also note, with that this formalism, we do not need to include any
term similar to the $k$--correction terms used in the computation of
standard luminosity functions, since the completeness curve by
definition allows to compute the maximum redshift used to estimate
$V_{\rm max}$.

We remark that the correction is computed for each galaxy at the
corresponding redshift, and that we will use here, at variance with
the rest of the paper, the ratio between the stellar mass $M_*$ and
the {\it observed} K--band flux $k$.  The correction is computed as
follows.

At a given mass $M^*_{\rm inf}
\leq M_* \leq M^*_{\rm compl}$, the observed K--band flux $k$
corresponding to a given stellar mass $M_*$ is encompassed between two
values, named $k_{\rm min}$ and $k_{\rm max}$ here (we have adopted in this
case of passively evolving, dust free objects). The key quantity is
the fraction $1-f_{\rm obs}$ of galaxies lost by effect of the incomplete
coverage of the $M_*/L_K$ ratio, that reside in  the shaded area of
Fig.~\ref{complet}: in practice, the correction is based on the
computation of the fraction of observed galaxies $f_{\rm obs}$, that will
be used to correct for the accessible volume element $V_{\rm max}$.

To show how we compute the fraction $f_{\rm obs}$, let us first define the
number density $N(M_*,k)$ of objects (at redshift $z$) with observed
$K_s$--band luminosity $k$ and mass $M$: in principle, the fraction
$f_{\rm obs}$ (at given $M$) can be estimated from the number density of objects
$N(M,k)$ as
\begin{equation}
f_{\rm obs} = {\int_{k_{\rm lim}}^{k_{\rm max}}{N(M,k) dk}
\over
\int_{k_{\rm min}}^{k_{\rm max}}{N(M,k) dk}}
\end{equation}

Since the shape of $N(M,k)$ at fluxes fainter than our $K=20$ limit is
in principle unknown, we have to assume that the distribution of the
\ml ratio {\it at fixed observed luminosities and at a given redshift}
is independent of the luminosity: although such a factorization
(i.e. the assumption that the distribution of the \ml ratio is
independent of the luminosity) does not likely hold at any $k$, we
actually need it to be valid around our $K=20$ limit, where we compute
our correction. Given the small range in luminosity that we sample
around the limit, we do not expect this assumption to invalidate the
computation of the correction.

In this case the number density $N(M,k)$ can be written as
\begin{equation}
N(M,k) = \mu(\frac{M}{k}) \phi(k)
\label{eq_dens}
\end{equation}
where $\mu(M/k)$ is the distribution of the \ml ratio, and $\phi(k)$
is the luminosity distribution of objects with given \ml.  $\phi(k)$
is linked to the luminosity distribution $\Phi(k)$ (i.e. to the galaxy
counts of objects at redshift $z$) by the requirement that
\begin{equation}
\Phi(k) = \int_{0}^{\infty}{ \phi(k) \mu(M/k)dM} \propto \phi(k) k
\label{eq_prop}
\end{equation}.

%A stronger and pippo assumption would be that the distribution
%$\mu(M/L)$ is constant (i.e.  independent also of \ml) within the
%physically allowable range: then $\phi(L) \propto \Phi(L)$, such that
%the fraction $f$ is directly the ratio of the relative counts due to
%galaxies at the corresponding redshift:

%\begin{equation}
%f ={ {10^{0.28\times 20}-10^{0.28\times K_{inf}}} \over
% {10^{0.28\times K_{sup}}-10^{0.28\times K_{inf}}}}
%\end{equation}
%where we take the slope of the K galaxy counts (dlogN/dK=0.28) from
%Saracco et al. 2000.

We have estimated $\Phi(k)$ on our data in the four redshift intervals
that we have used, and found that (at $18<K<20$) it can be well
represented by a power law :
\begin{equation}
\Phi(k)\propto k^{-2.5\alpha_z-1}
\label{eq_phi}
\end{equation}
with a redshift-dependent index $\alpha_z=0.2,0.2,.0.33,0.38$ at
$z=0.45,0.9,1.3,1.75$, that is consistent with the overall slope of
the counts (dlogN/dK=0.28) from Saracco et al. 2001.

The simplest approach would be to assume that the distribution function
 $\mu(M/k)$ is constant at fixed $k$ : with this very coarse
 assumption the correction factor $f_{\rm obs}$ becomes

\begin{equation}
f_{\rm obs} (z,M)=
{\int\limits_{k_{\rm lim}}^{k_{\rm max}}{ k^{-2.5\alpha_z-2}dk}
\over
\int\limits_{k_{\rm min}}^{k_{\rm max}}{ k^{-2.5\alpha_z-2}dk}
}
=
{k_{\rm max}^{-2.5\alpha_z-1}-k_{\rm lim}^{-2.5\alpha_z-1}
\over
k_{\rm max}^{-2.5\alpha_z-1}-k_{\rm min}^{-2.5\alpha_z-1}}
\end{equation}
.

Alternatively, one can explicitly take into account the intrinsic
distribution in the \ml ratio.  We have already made the assumption
that the distribution $\mu(M/k)$ is independent of $k$, at a given
$z$, which implies that at the same $z$ the distribution of $M/L$ is
independent of $L$, where L is the rest frame luminosity in the
wavelength range corresponding to the redshifted K band. We will
further assume that the distribution of $M/L$ is also constant {\it
within each redshift bin}, and that within the redshift bin we can
ignore the differential $k$--correction (that are indeed small in the
K band (Pozzetti et al. 2003). That is, we assume that one can write
$L=D_L(z)^2 k$, where $D_L(z)$ is the corresponding luminosity distance.

With this assumptions, we have found that 
the \ml distribution function can be conveniently
expressed as 
\begin{equation}
\mu(M/L)\propto(\frac{M}{L})^\gamma e^{- \frac{1}{\mu} \frac{M}{L}}
\label{eq_mu}
\end{equation}

such that the distribution of the observed $M/f$ becomes:

\begin{equation}
\mu(M/k)\propto(\frac{M}{k})^\gamma e^{- \frac{1}{\mu(z_c)} \frac{D_L^2(z_c)}{D_L^2(z)}\frac{M}{k}}
\label{eq_mu}
\end{equation}

where $z_c$ is the center of the redshift bin.

We have found that with $\gamma=n-2.5\alpha_z$ (with $n$ integer)
this expression provides an excellent fit to the \ml
distribution at $19<K<20$ and, most important, makes the fraction
$f$ (Eq. 1) analytical: indeed, substituting Eq.  \ref{eq_phi} into
Eq. \ref{eq_prop}, and then inserting Eq. \ref{eq_prop} and
\ref{eq_mu} into Eq.  \ref{eq_dens}, the fraction of observed objects
$f_{\rm obs}$ becomes:
\begin{equation}
f_{\rm obs}(z,M) =
{\int\limits_{k_{\rm lim}}^{k_{\rm max}}{ k^{-n-2}e^{-\frac{1}{\mu(z_c)}\frac{D_L^2(z_c)}{D_L^2(z)}\frac{M}{k}}dk}
\over
\int\limits_{k_{\rm min}}^{k_{\rm max}}{ k^{-n-2}e^{-\frac{1}{\mu(z_c)}\frac{D_L^2(z_c)}{D_L^2(z)}\frac{M}{k}}dk}
}
=
\end{equation}
\begin{equation}
=
{
\Gamma(n+1,{D_L^2(z_c) \over \mu(z_c) D_L^2(z)}{M\over k_{\rm lim}}) - \Gamma(n+1,{D_L^2(z_c) \over \mu(z_c) D_L^2(z)}{M\over k_{\rm max}}) 
\over
\Gamma(n+1,{D_L^2(z_c) \over \mu(z_c) D_L^2(z)}{M\over k_{\rm min}}) - \Gamma(n+1,{D_L^2(z_c) \over \mu(z_c) D_L^2(z)}{M\over k_{\rm max}}) 
}
\end{equation}

The best fit values for $n$ and $\mu(z_c)$ have been found at each
redshift bin to be: $n=4,4,5,6$ and $\mu = 0.08,0.5,1.2,2.4$ (when
masses are in units of $10^9 M_\odot$ and fluxes are in units of
$10^{-29}$erg/cm$^2$/s/Hz) at $z_c=0.45,0.9,1.3,1.75$, respectively.
As stated before, the correction factor $f_{\rm obs}$ is then multiplied
to the volume element $V_{\rm max}$ of any galaxy with mass $M^*_{\rm inf}
\leq M_* \leq M^*_{\rm compl}$, both in the $1/V_{\rm max}$ binned GSMF as
well as in the Schechter best-fit. The correction is applied until it
exceeds a factor of two. The selection curves shown in
Fig.~\ref{mass_z} have been computed with this criteria.
\begin{figure}
\centering
\resizebox{\hsize}{!}{\includegraphics{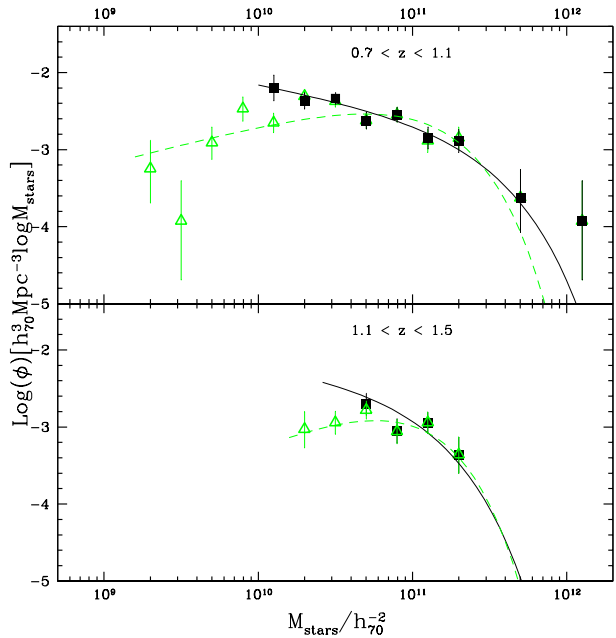}}
\caption{ Effect of the correction for incompleteness on the Galaxy
Stellar Mass Function in two redshift bins. Empty circles represent
the GSMF computed on strictly mass--complete samples, while
filled squares represent the GSMF
computed on the ``extended'' sample with the correction for
incompleteness applied, and solid line is the corresponding 
Schechter fit.  Triangles show the GSMF computed on the
``extended'' sample without any correction for incompleteness, and
dashed line the corresponding Schechter fit.  }
\label{conf_comp}
\end{figure}

The practical effects of this correction are shown in
Fig.~\ref{conf_comp}, where we compare the GSMF with and without the
applied correction, where we show that if we entirely ignore the
incompleteness effects, the GSMF appears to drop significantly in the
low mass bins. 
% When we compute the GSMF in the ``strict''
%mass--complete sample, the effect is to reduce the depth where the
%analysis can be carried on, and to increase the size of the error bars
%on the low mass bins.  The incompleteness--corrected GSMF is instead
%able to recover most of the sample and to reduce the statistical
%noise, while keeping a faint slope consistent with the ``strictly''
%complete one.  Indeed we have found that the Schechter parameters
%derived by using the ``strictly complete'' or the ``extended''
%selection criteria are consistent with each other.

\end{document}